\documentclass[review]{elsarticle}
\usepackage{lineno,hyperref}
\usepackage{graphicx}
\usepackage{ulem}
\usepackage{color}
\usepackage{amsmath}
\usepackage{mathabx}
\usepackage{listings}
\newcommand{\eqnref}[1]  {equation~(\ref{#1})}
\newcommand{\figref}[1]  {Fig.~\ref{#1}}
\newcommand{\eqnrefs}[1] {equations~(\ref{#1})}
\newcommand{\secref}[1]  {Section~\ref{#1}}

\newcommand{\Secref}[1]  {Section~\ref{#1}}

\newcommand{\smg}[1]    {\textcolor{blue}{#1}}

\usepackage[font=small,skip=0pt]{caption}
\modulolinenumbers[5]

\journal{Journal of Mathematical Psychology}

\begin{document}

\begin{frontmatter}

\title{The free energy principle for action and perception: A mathematical review}

\author[mymainaddress]{Christopher L. Buckley\corref{mycorrespondingauthor}\fnref{myfootnote}}
\author[mysecondaryaddress]{Chang Sub Kim\fnref{myfootnote}}
\author[mymainaddress]{Simon McGregor}
\author[mymainaddress]{Anil K. Seth}

\fntext[myfootnote]{Joint contribution}
\cortext[mycorrespondingauthor]{Corresponding author}
\address[mymainaddress]{School of Engineering and Informatics, Evolutionary and Adaptive Systems Group, University of Sussex, Brighton, BN1 9QJ, UK}
\address[mysecondaryaddress]{Department of Physics, Chonnam National University, Gwangju 61186, Republic of Korea}

\begin{abstract}
The `free energy principle' (FEP) has been suggested to provide a unified theory of the brain, integrating data and theory relating to action, perception, and learning. The theory and implementation of the FEP combines insights from Helmholtzian `perception as inference', machine learning theory, and statistical thermodynamics.  Here, we provide a detailed mathematical evaluation of a suggested biologically plausible implementation of the FEP that has been widely used to develop the theory. Our objectives are (i) to describe within a single article the mathematical structure of this implementation of the FEP; (ii) provide a simple but complete agent-based model utilising the FEP; (iii) disclose the assumption structure of this implementation of the FEP to help elucidate its significance for the brain sciences.
\end{abstract}


\end{frontmatter}

\linenumbers
\section{Introduction}
\label{Introduction}
The brain sciences have long searched for a `unified brain theory' capable of integrating experimental data relating to, and disclosing the relationships among, action, perception, and learning.  One promising candidate theory that has emerged over recent years is  the `free energy principle' (FEP) \cite{Friston2010Nature,Friston2009Trend}. The FEP is ambitious in scope and attempts to extend even beyond the brain science to account for adaptive biological processes spanning an enormous range of time scales, from millisecond neuronal dynamics to the tens of millions of years span covered by evolutionary theory \cite{Friston2010Trends,Friston2010Nature}.

The FEP has an extensive historical pedigree. Some see its origins starting with Helmholtz' proposal that perceptions are extracted from sensory data by probabilistic modelling of their causes \cite{Helmholtz}.  Helmholtz also originated the notion of thermodynamic free energy, providing a second key inspiration for the FEP \footnote{Thermodynamic free energy describes the macroscopic properties of nature, typically in thermal equilibrium where it takes minimum values,
in terms of a few tractable variables.}. These ideas have reached recent prominence in the `Bayesian brain' and `predictive coding' models, according to which perceptions are the results of Bayesian inversion of a causal model, and causal models are updated by processing of sensory signals according to Bayes' rule \cite{knill:2004,rao:1999,bubic:2010,clark2013whatever}. However, the FEP  naturally accommodate and description of both action and perception within the same framework \cite{Friston2010BC} thus other  see it's  origins  in 20th-century cybernetic principles of homeostasis and predictive control \cite{seth2015cybernetic}. 

A recognisable precursor to the FEP as applied to brain operation was  developed by Hinton and colleagues, who showed that a function resembling free energy could be used to implement a variation of the expectation-maximization algorithm \cite{NealHinton:1998}, as well as for training autoencoders \cite{HintonZemel:1994} and learning population codes \cite{Zemel:1995}. Because these algorithms integrated Bayesian ideas with a notion of free energy, Hinton named them as `Helmholtz machines' \cite{dayan:1995}. The FEP builds on these insights to provide a global unified theory of cognition. Essentially, this work generalizes these  results by noting that all (viable) biological organisms resist a tendency to disorder as shown by their homeostatic properties (or, more generally, their autopoietic properties), and must therefore minimize the occurrence of events which are atypical (`surprising') in their habitable environment. For example, successful fish typically find themselves surrounded by water, and very atypically find themselves out of water, since being out of water for an extended time will lead to a breakdown of homeostatic (autopoietic) relations. Because the distribution of `surprising' events is in general unknown and unknowable, organisms must instead minimise a tractable proxy, which according to the FEP turns out to be `free energy'. Free energy in this context is an information-theoretic construct that (i) provides an upper bound on the extent to which sensory data is atypical (`surprising') and (ii) can be evaluated by an organsim, because it depends eventually only on sensory input and an internal model of the environmental causes of sensory input.  While at its most general this theory can arguably be applied to all life-processes \cite{friston2013life}, it provides a particularly appealing  account of brain function. Specifically it describes how neuronal processes could implement free energy minimisation either by changing sensory input via action on the world, or by updating internal models via perception, with implications for understanding the dynamics of, and interactions among action, perception, and learning. These arguments have been developed in a series of papers which have appeared over the course of the last several years \cite{Friston2005Phil,Friston2006Physio,Friston2006NeuroImage, Friston2007Synthe,Friston2008NeuroImage,Friston2008One,Friston2009One, Friston2009Neural,Friston2010BC,Friston2009PhilTrans,Friston2010Brain,adams2013predictions,friston2010generalised,pezzulo2015active,friston2016active}.

The FEP deserves close examination because of the  claims made for its explanatory power. It has been suggested that the FEP discloses novel and straightforward relationships among fundamental psychological concepts such as memory, attention, value, reinforcement, and salience \cite{Friston2009Trend}. Even more generally, the FEP is claimed to provide a ``mathematical specification of `what' the brain is doing'' [\cite{Friston2009Trend}, p.300], to unify perception and action \cite{Friston2010BC}, and to provide a basis for integrating several general brain theories including the Bayesian brain hypothesis, neural Darwinism, Hebbian cell assembly theory, and optimal control and game theory \cite{Friston2010Nature}. The FEP has even been suggested to underlie Freudian constructs in psychoanalysis \cite{Friston2010Brain}.

Our purpose here is first to supply a mathematical appraisal of the FEP, which we hope will facilitate evaluation of claims such as those listed above; note that we do \textit{not} attempt to resolve any such claims here. A mathematical appraisal is worthwhile because the FEP combines advanced concepts from several fields, particularly statistical physics, probability theory, machine learning, and theoretical neuroscience.  The mathematics involved is non-trivial and has been presented over different stages of evolution and using varying notations. Here we first provide a  complete technical account of the FEP, based on a history of publications through which the framework has been developed. Second we provide a complete description of simple agent based model working under this formulation.  While we note that several other agent based models have been presented they have often made use of existing toolboxes which, while powerful, have perhaps clouded a fuller understanding of the FEP. Lastly we use our account to identify the assumption structure of the FEP, highlighting several instances in which non-obvious assumptions are required. 

In the next section we provide a brief overview of the FEP followed by detailed guide to the technical content covered in the rest of the paper.

\section{An overview of the FEP}
\label{overview}
Broadly the FEP is an account of cognition  derived from the consideration of  how biological organisms maintain their state away from thermodynamic equilibrium with their ambient surroundings. The argument runs that organisms are mandated, by the very fact of their existence, to minimize the dispersion of their constituent states.  The atypicality of an event can be quantified by the negative logarithm of the probability of its sensory data, which is commonly known in information theory as `surprise' or `self-information' and the overall atypicality of an organism's exchanges with its environment can be quantified as a total lifetime surprise \cite{Friston2010Nature,Friston2009Trend}. The term surprise has caused much confusion since it is distinct from the subjective psychological phenomenon of surprise. Instead, it is a measure of how atypical a sensory exchange is.  This kind of surprise can be quantified using the standard information-theoretic log-probability measure
\[ -\ln p(\varphi) \] where $p(\varphi)$ is the probability of observing some particular sensory data
$\varphi$ in a typical (habitable) environment. Straightforwardly this quantity is large if the probability of the observed data is small and zero if  the data is fully expected, i.e., probability  $1$. To avoid  confusion with the common-sense meaning of the word `surprise' we will refer to it as ``surprisal'' or ``sensory surprisal''.

\subsection{R- and G- Densities}
The FEP argues organisms cannot minimise surprisal directly, but instead minimise an upper bound called `free energy'. To achieve this it is proposed that all (well adapted) biological organisms maintain a probabilistic model of their typical (habitable) environment (which includes their body), and attempt to minimize the occurrence of events which are atypical in such an environment as measured by this model.   Two key probability densities are necessary to evaluate free energy.  First it is suggested that organisms maintain an implicit account of a best guess at the relevant variables that comprise their environment (i.e. those variables which cause its sensory data). This account is in the form of a probability distribution over all possible values of those variables, like a Bayesian belief; this model is instantiated, and parameterised, by physical variables in the organism's brain such as neuronal activity and synaptic strengths, respectively.  When an organism receives sensory signals, it  updates this distribution to better reflect the world around it, allowing it to effectively model its environment. In other words, the organism engages in a process similar to Bayesian inference regarding the state of its environment, based on sensory observations. This internal model of environmental states is called the  ``recognition density'' or the R-density. In order to update the R-density appropriately, the organism needs some implicit assumptions about how different environmental states shape sensory input. These assumptions are presumed to be in the form of a joint probability density between sensory data and environmental variables, the ``generative density'',  or G-density. This density is also presumed to be encoded within the organsims brain. As we will see, following a Bayesian formalism, this joint density is calculated as the product of two densities; a \textit{likelihood} describing the probability of sensory input given some environmental state and a \textit{prior} describing the organisms current "beliefs" of the probability distribution over environmental states.

\subsection{Minimising Free Energy}
Free energy  is a (non-negative) quantity formed from the Kullback-Leibler divergence between the R- and G-densities. Consequently, it is not a directly measurable physical quantity: it depends on an interpretation of brain variables as encoding notional probability densities. Note: the quantity 'free energy' is distinct from thermodynamic free energy thus here we will refer to it as  \textit{informational free energy} (IFE).

Minimisation of IFE has two functional consequences. First it provides an upper bound on sensory surprisal. This allows organisms to estimate the dispersion of their constituent states and is central to the interpretation of FEP as an account of life processes \cite{Friston2010Nature}.  However, IFE minimisation also plays a central role in a Bayesian approximation method. Specifically  ideal (exact) Bayesian inference, in general, involves evaluating difficult integrals and thus a core hypothesis of the FEP framework is that the brain implements approximate Bayesian inference in an analogous way to a method known as variational Bayes. It can be shown that minimising IFE makes the R-density a good approximation to posterior density of environmental variables given sensory data.  Under this interpretation the surprisal term in the IFE becomes more akin to the negative of \textit{log model evidence} defined in more standard implementations of variational Bayes \cite{hinton1994autoencoders}.

\subsection{The Action-Perception Cycle}
Minimising IFE by updating the R-density provides an upper-bound on surprisal but cannot minimise it directly. The FEP suggests that organisms also act on their environment to change sensory input, and thus minimise surprisal indirectly \cite{Friston2010Nature,Friston2009Trend}. The mechanism underlying this process is formally symmetric to perceptual inference, i.e., rather than inferring the cause of sensory data an organism must infer actions that best make sensory data accord with an internal environmental model \cite{Friston2010BC}. Thus, the mechanism is often referred to as \textit{active inference}.  Formally, action allows an organisms to avoid the dispersion of its constituent states and is suggested to underpin a form of, homoeostasis, or perhaps more precisely homeorhesis \cite{seth2015cybernetic}. However, equivalently, one can view action as satisfying  hard constraints encoded in the organisms environmental model \cite{Friston2010BC}. Here expectations in the organism's G-density (its "beliefs" about the world) cannot be met directly by perception and thus an organism must act to satisfy them. In effect these expectations effectively encode the organism's \textit{desires} on environmental dynamics. For example, the organisms model may prescribe it maintains a desired local temperature; we will see an example of this in \secref{action}. Here action is seen as  more akin to control \cite{seth2015cybernetic} where behaviour arises from a process of minimising deviations between the organisms actual and a desired trajectory \cite{Friston2010BC}. Note: an implicit assumption here is that these constraints are conducive to the organisms survival \cite{Friston2010Nature,Friston2009Trend}, perhaps arrived at by an evolutionary process. Other different roles for action within the FEP have also been suggested, e.g., action as a process of experimentation with the goal to disambiguate competing environmental models \cite{friston2012perceptions,seth2015cybernetic}. However, here we only consider action as a source of control \cite{Friston2010BC,seth2015cybernetic}.

\subsection{Predictive Coding}
There at least two general ways to view most FEP-based research. First the central theory \cite{Friston2006Physio} which offers a particular explanation of cognition in terms of Bayesian inference. Second  a biologically plausible \textit{process theory} of how the relevant probability densities could be parameterised by variables in the brain (i.e. a model of what it is that brain variables encode), and how the variables should be expected to change in order to minimize IFE. The most commonly used implementation  of the FEP, and the one we focus on here,  is  strongly analogous with the predictive coding framework \cite{rao:1999}. Specifically predictive coding theory constitutes one plausible mechanism whereby an organism could update its environmental model (R-density) given a belief of how its environment works (G-density).  The concept of predictive coding  overturns classical notions of perception (and cognition) as a largely bottom-up process of evidence accumulation or feature detection driven by impinging sensory signals, proposing instead that perceptual content is determined by top-down predictive signals arising from multi-level generative models of the environmental causes of sensory signals, which are continually modified by bottom-up prediction error signals communicating mismatches between predicted and actual signals across hierarchical levels (see \cite{clark2013whatever} for a nice review). In the context of the FEP the R-density is updated using a hierarchical predictive coding (see \secref{learning}). This has several theoretical benefits. Firstly, under suitable assumption IFE becomes formally equivalent to prediction error (weighted by confidence terms), which can readily be computed in neural hardware. Hierarchical coding also provides a very generic prior which allows high-level abstract sensory features to be learned from the data, in a manner similar to deep learning nets \cite{hinton2007learning}. Finally, the sense in which the brain models the environment can be conceptualised in a very direct way as the prediction of sensory signals. We will also  see in \secref{learning} that this implementation suggests that we do not even need to know what environmental features the R- and G-densities constitute a model of. Given appropriate assumptions, the formalism can be rewritten to depend only on predictions of sensory data, along with recursive predictions of the brain variables which encode those predictions. 

\subsection{A technical guide}

In the rest of this work we review the FEP in  detail but first we provide a detailed guide to each section. Most of what we present is related to standard concepts and techniques in statistical mechanics and machine learning. However, here we present these ideas in detail to make clear their role for the  FEP as theory of biological systems.

In \Secref{IFE} we describe the core technical concepts of  FEP including the R-density,  G- density, and  IFE.   We  show how minimising IFE has two consequences. First, it makes the  R-density a better estimate of posterior beliefs about environmental state given sensory data, thus implementing approximate Bayesian inference. Second, it makes the IFE itself an upper-bound on sensory surprisal.

In \secref{laplaceencoding}  we discuss the approximations that allow the brain to explicitly instantiate the R-density and thus specify IFE. Specifically, we make the approximation that the R-density take Gaussian form, the \textit{Laplace approximation}, and that brain states, e.g. neural activity, represent the sufficient statistics of this distribution (mean and variance). Utilising this form for the R-density and various other approximations we derive an expression for the IFE in terms of the unknown G-density only; we refer to this approximation as the \textit{Laplace encoded energy}. The derivations in this section are done for the univariate Gaussian case, but we  give an expression for the full multivariate case at the end of the section.

In \secref{inference} we  look at different forms for the G-density. We start by specifying simple generative models which comprise the brain's  model of how the world works, i.e., how sensory data is caused by environmental (including bodily) variables. We utilise these generative models to specify the brain's expectation on environmental states given sensory data in terms of a Gaussian distribution parametrised by expected means and variances (inverse precisions) on brain states. Combining this with the result of the last section allows us to write an  expression for  Laplace encoded-energy as a quadratic sum of  prediction errors (difference between expected and actual brain states given sensory data) modulated by expected variances (or inverse precisions), in line with predictive-coding process theories. Initially we show this for a static generative model but extend it to include dynamic generative models by introducing the concept of generalised motion. Again we derive the results for the univariate case but provide expressions for the multivariate case.
 
In \secref{minimisation} we show how the brain could dynamically minimises IFE. Specifically, we describe how brain states are optimised to minimise IFE through gradient descent. We discuss complications of this method when considering dynamical generative models.

\Secref{action} demonstrates how action can be implemented as a similar gradient descent scheme. Specifically we show how, given a suitable \textit{inverse model}, actions are chosen to change sensation such that they minimise IFE. We ground this idea, and the mechanisms for perception described in prior sections, in a simple agent based simulation. We show how an agent with an appropriate model of the environment, can combine action and perception to minimise IFE constrained both by the environment and its own expectations on brain states.

In \secref{learning} we extend the formalism to include learning. Specifically we show how the brain could modify and learn the G-density. To facilitate this we describe notion of hierarchical generative models which involve empirical priors. We lastly describe a gradient descent schemes which allows the brain to infer parameters and hyperparameters of the IFE and thus allow the brain to learn environmental dynamics based on sensory data.

Finally, \secref{discussion} summarizes the FEP and discusses the implications of its assumption structure for the brain sciences.

\section{Informational free energy}
\label{IFE}

\begin{table}[b]
\begin{tabular}{lp{1.7in}p{3.2in}}
\multicolumn{3}{l} {\bf Table 1. Mathematical Objects in the IFE}\\
\hline \hline \multicolumn{1}{l} {\bf Symbol} & \multicolumn{1}{l}
{\bf Name} & \multicolumn{1}{l} {\bf Description}\\ \hline

& & \\
$\vartheta$ & Environmental states & These refer to all states outside of the brain and include both environmental and bodily variables.  \\

$\varphi$ & Sensory data & Signals caused by the environment. \\

$q(\vartheta)$ & R-density & Organism's (implicit) probabilistic representation of environmental states which cause sensory data. \\

$p(\varphi,\vartheta)$ & G-density & Joint probability density, encoded in the brain relating sensory data to environmental states. Assumed to be encoded in a form which makes $p(\varphi|\vartheta)$ and $p(\vartheta)$ accessible, but not $p(\vartheta|\varphi)$ or $p(\varphi)$. \\

$p(\vartheta)$ & Prior density & Organism's prior beliefs, encoded in the brain's state, about environmental states. \\

$p(\varphi|\vartheta)$ & Likelihood density & Organism's implicit beliefs about how environmental states map to sensory data. \\

$p(\vartheta|\varphi)$ & Posterior density & The inference that a perfectly rational agent (with incomplete knowledge) would make about the environment's state upon observing new sensory information, given the organism's prior assumptions. \\

$p(\varphi)$ & Sensory density & Probability density of the sensory input, encoded in the brain's state, which cannot be directly quantified given sensory data alone. \\

$-\ln p(\varphi)$ & Surprisal & \textit{Surprise} or \textit{self-information} in
information-theory terminology, which is equal to the negative of \textit{log model evidence} in Bayesian statistics. \\

$F$ & Information-theoretic free energy (IFE) &  The quantity minimised under the FEP which forms an upper bound on surprisal allows the approximation of the posterior density. \\

\hline

\end{tabular}
\end{table}

We start by considering a world that consists of  a brain and its surrounding body/environment. For the rest of the presentation we refer to the  body and environment as simply the \textit{environment} and use this to refer to all processes outside of the brain. The brain is distinguished from its environment by an interface which is not necessarily a physical boundary but rather may be defined functionally; thus the boundary could reside at the sensory and motor surfaces rather than, for example, at the limits of the cranial cavity. The environment is characterized by states, denoted collectively as $\{\vartheta\}$, which include well-defined characteristics like temperature or the orientation of a joint  but also unknown and uncontrollable states, all evolving according to physical laws. The environmental states, as exogenous stimuli, give rise to sensory inputs for which the symbols $\{\varphi\}$ are designated collectively. These sensory inputs are assumed to reside at the functional interface distinguishing the brain from the environment, and we assume a many-to-one (non-bijective) mapping between $\{\vartheta$\} and $\{\varphi\}$ \cite{friston:nrnc2:2010}. We further assume that the brain, in conjunction with the body, can perform actions to modify sensory signals.
	
We assume that the important states of the environment cannot be directly perceived by an organism but instead must be inferred by a process of Bayesian inference.  Specifically, the  goal of the agent is to determine the probability of  environmental states given its sensory input.  To achieve this we assume organism's encodes prior beliefs about  these states characterized by the joint density $p(\vartheta,\varphi)$ or  G-density. Where the G-density can be factorized into (with respect to $\vartheta$), the \textit{prior} $p(\vartheta)$ (corresponding to the organism's "beliefs" about the world before sensory input is received) and a \textit{likelihood} $p(\varphi| \vartheta)$ (corresponding to the organism's assumptions about how environmental dynamics cause sensory input), 
\begin{equation}
\label{gdensity}
p(\vartheta,\varphi) = p(\varphi|\vartheta)p(\vartheta).
\end{equation}
Give an observation,  $\varphi=\phi$ (e.g. some particular sensory data), a \textit{posterior} belief about the environment can then be written as  $p(\vartheta|\varphi=\phi)$.   This quantity can be calculated  using the  prior and likelihood using Bayes theorem as,

\begin{equation}\label{Bayes}
p(\vartheta|\phi)
= \frac{1}{p(\varphi=\phi)}p(\phi|\vartheta)p(\vartheta)=
\frac{p(\phi|\vartheta)p(\vartheta)}{\int p(\phi|\vartheta) p(\vartheta) d\vartheta}.
\end{equation}
All the probability densities are assumed to be normalized as
\[
\int d\vartheta \int d\varphi~ p(\vartheta,\varphi)=\int d\vartheta~p(\vartheta)
= \int d\varphi~p(\varphi)=1,
\]
where $p(\vartheta)$ and $p(\varphi)$ are the reduced or marginal probability-densities conforming to
\begin{equation} \label{marginal}
p(\vartheta)=\int d\varphi~ p(\vartheta,\varphi) \quad
{\rm and}\quad p(\varphi)=\int d\vartheta ~ p(\vartheta,\varphi).
\end{equation}

To calculate the posterior probability it is necessary to evaluate  the marginal integral, $\int p(\phi|\vartheta) p(\vartheta)\vartheta$, in the denominator of equation~(\ref{Bayes}). However, this is often difficult, if not practically intractable. For example,  when continuous functions are used to approximate the likelihood and prior,  the integral may be analytically intractable. Or in the discrete case, when this integral reduces to a sum, the number of calculations may grow exponentially with the number of states. \textit{Variational Bayes} (sometimes known as `ensemble learning')  is a method for (approximately) determining $p(\vartheta|\varphi)$ which avoids the evaluation of this integral, by introducing an optimization problem \cite{Friston2008NeuroImage}. Such an approach requires an auxiliary probability density, representing the current `best guess' of the causes of  sensory input. This is the \textit{recognition density}, or R-density, introduced in the overview.
Again the R-density is also normalised as:
\begin{equation}\label{norm1}
\int q(\vartheta) d\vartheta = 1.
\end{equation}
We can construct a measure of the difference between this density and the true posterior in terms of an information-theoretic divergence, e.g., the Kullback-Leibler divergence \cite{Cover}, i.e.,
 \begin{equation}
\label{KL}
D_{KL}(q(\vartheta) ||p(\vartheta|\varphi))    = \int d\vartheta~ q(\vartheta)\ln \frac{q(\vartheta)}{p(\vartheta|\varphi)} 
\end{equation}
An  R-density that minimises this divergence would provide a good approximation to the true posterior. But obviously we cannot evaluate this quantity because we still do not know the true posterior. However, we can rewrite this equation  as, 
\begin{equation}
\label{KLIFE}
 D_{KL}(q(\vartheta) ||p(\vartheta|\varphi))  =  F + \ln p(\varphi)
\end{equation}
where we have defined F as the \textit{informational free energy} (IFE),
 \begin{equation}\label{FFE} 
 F \equiv \int d\vartheta~ q(\vartheta)\ln \frac{q(\vartheta)}{p(\vartheta,\varphi)} .
 \end{equation} 
Note here we have introduced the G-density to the denominator on the right-hand side. In contrast to \eqnref{KL} we can evaluate IFE directly because it depends only on the R-density, which we are free to specify, and the G-density, i.e., a model of the environmental causes of sensory input. Furthermore, the second term on the right-hand side in \eqnref{KLIFE} only depends on sensory input and is independent of the form of the R-density. Thus, minimising \eqnref{FFE} with respect to the R-density will also minimise the Kullback-Leibler divergence between the R-density and the true posterior. Thus, the result of this minimisation will make the R-density approximate the true posterior.

The minimisation of IFE also suggests an indirect way to estimate surprisal. Specifically according to Jensen's inequality \cite{Cover}, the  Kullback-Leibler divergence is always greater than zero. This implies the inequality,
\begin{equation}\label{surprise}
F\ge  - \ln p(\varphi).
\end{equation}
which means that the IFE also provides an upper bound on the \textit{surprisal} as described in \secref{Introduction}.  However, note the IFE is equal to surprisal only when the R-density $q(\vartheta)$ becomes identical with the posterior density $p(\vartheta|\varphi)$; i.e., it is this condition that specifies when IFE provides a \textit{tight bound} on surprisal (see \secref{overview}).  Furthermore, while this process furnishes the organism with  an approximation of  surprisal it does not minimise it. Instead the organism can minimise IFE further  by minimising surprisal indirectly by acting on the environment and changing sensory input, see \secref{action}. 

 Note: formally $p(\varphi)$, which describes the agent's internal (implicit) probabilistic predictions of sensory inputs, should be written as as $p(\varphi | m)$. This follows a convention in Bayesian statistics to indicate that a reasoner must begin with some arbitrary prior before it can learn anything; $p(\varphi)$ indicates the prior assigned to $p$ ab initio by agent $m$. However,  this notation is unwieldly and does not change the derivations that follow thus we will omit this for the rest of the presentation.

There are several analogies between the terms in the formalism above and the formulation of Helmoltz' thermodynamic free energy.  These terms can serve as useful substitutions in the derivation to come and, thus, we describe them here. Specifically when the G-density is unpacked in \eqnref{FFE}, the IFE splits into two terms, 
\begin{equation}\label{Helmholtz}
 F = \int d\vartheta~ q(\vartheta)E(\vartheta,\varphi) + \int d\vartheta~ q(\vartheta)\ln q(\vartheta) 
 \end{equation}
where, formally speaking, the first term in equation~(\ref{Helmholtz}) is an average of the quantity
\begin{equation}\label{energy}
E(\vartheta,\varphi)\equiv -\ln p(\vartheta,\varphi)
\end{equation}
over the R-density $q(\vartheta)$ and the second term is essentially the negative entropy associated with the recognition density. By analogy with Helmoltz' thermodynamic free energy the first term in \eqnref{Helmholtz} is called \textit{average energy} [Accordingly, $E(\vartheta,\varphi)$ itself may be termed the \textit{energy}] and the second term the negative of \textit{entropy} \cite{Thermodynamics}.

In summary, minimising IFE with respect to the R-density, given an appropriate model for the G-density  $p(\vartheta,\varphi)$ in which the sensory inputs are encapsulated, allows one to approximate the Bayesian posterior. Furthermore minimising IFE through perception  also gives a lower bound on the sensory surprisal. 

Table 1 provides a summary of the mathematical objects associated with the IFE.

\begin{table}[bh!]
\begin{tabular}{lp{1.6in}p{3.0in}}
\multicolumn{3}{l} {\bf Table 2. Mathematical objects in the Laplace encoding}\\
\hline \hline \multicolumn{1}{l} {\bf Symbol} & \multicolumn{1}{l}
{\bf Name} & \multicolumn{1}{l} {\bf Description}\\ \hline
& & \\
$F[q(\vartheta);\varphi]$ & Variational IFE  & A functional (higher-order function) of the R-density $q(\vartheta)$
and a function of the sensory data $\varphi$. \\

$\mathcal{N}(\vartheta;\mu,\zeta)$ & (Gaussian) fixed-form R-density & An `ansatz' for
 unknown $q(\vartheta)$ (the Laplace approximation) \\

$\mu$, $\zeta$ & Parameters for the R-density & Sufficient statistics
(expectation and variance) of the fixed-form R-density, encoded in the brain's state. \\

$\zeta^*$ & Optimal variance & Analytically derivable optimal $\zeta$,
removing an explicit dependence of $F$ on $\zeta$. \\

$p(\varphi,\mu)$ & Laplace-encoded G-density & The G-density in which dependence
on $\vartheta$ has been replaced with a dependence on $\mu$. \\

${E}(\mu,\varphi)$ & Laplace-encoded energy & Mathematical construct defined
to be $-\ln p(\mu,\varphi)$. \\
\hline

\end{tabular}
\end{table}
\section{The  R-density: How the brain encodes environmental states}
\label{laplaceencoding}

To implement the method described above the brain must explicitly encode the R-density. To achieve this it is suggested that neuronal quantities (e.g., neural activity) parametrize  \textit{sufficient statistics} (e.g., means and variances, see later) of a probability distribution. More precisely the  neuronal variables encode a family of probability densities over environmental states, $\vartheta$. The instantaneous state of the brain $\mu$ then picks out a particular density $q(\vartheta; \mu)$ (the R-density) from this family; the semicolon in $q(\vartheta; \mu)$ indicates that $\mu$ is a parameter rather than a random variable. 

Finding the optimal $q(\vartheta; \mu)$ that minimises IFE in the most general case is intractable and thus further approximations about the form of this density are required. Two types of approximation are often utilised. First, an assumption that the R-density $q(\vartheta)$ can be factorised into independent sub-densities $q_1(\theta_1) \times \cdots q_N(\theta_N)$. Under this assumption the optimal R-density still cannot be expressed in closed form  but an approximate solution (of general form) can be improved iteratively \cite{Friston2008NeuroImageTECH}. This leads to a formal solution in which the sub-densities affect each other only through mean-field quantities. Approaches that utilise this  form of the R-density are often referred to an \textit{ensemble learning}. This approach is not the focus of the work presented here but for completeness we provide a treatment of unconstrained ensemble learning in \ref{EnsembleLearning}.

A more common approximation is to assume that the R-density take  Gaussian form, the so called \textit{Laplace approximation} \cite{Friston2008NeuroImage}. In this scenario, the sufficient statistics of the Gaussian become parameters which can be optimized numerically to minimize IFE.  For example the R-densities take the form
\begin{equation}
\label{Gaussian1}
q(\vartheta) \equiv \mathcal{N}(\vartheta;\mu,\zeta)
=\frac{1}{\sqrt{2\pi\zeta}} \exp\left\{
-(\vartheta-\mu)^2/(2 \zeta) \right\}
\end{equation}
where $\mu$ and $\zeta$ are the mean and variance values of a single environmental variable $\vartheta$.
Substituting this form for the R-density  into \eqnref{FFE}, and carrying out the integration produces a vastly simplified expression for the IFE. In following we examine this derivation in detail. For the clarity of  presentation we pursue it in the univariate case which captures all the relevant assumptions for the multivariate case.  We write the formulation for the multivariate case at the end of the section. For notational ease we define
\begin{equation} Z \equiv {\sqrt{2\pi\zeta}} \quad {\rm and}\quad {\cal E}(\vartheta)
\equiv (\vartheta-\mu)^2/(2 \zeta),
\label{PF}
\end{equation}
to  arrive at
\begin{equation}
q(\vartheta; \mu, \zeta)  = \frac{1}{Z}e^{-{\cal E}(\vartheta)},
\label{total}
\end{equation}
where  here we have drawn on  terminology from statistical physics in which the normalization factor $Z$ is called the \textit{partition function} and ${\cal E}(\vartheta)$ the energy of the subsystem $\{\vartheta\}$ \cite{SM}. Substituting  this equation  into \eqnref{Helmholtz} and carrying out the integration  leads to a much simplified expression for IFE :
\begin{eqnarray}
F &=& \int d\vartheta~q(\vartheta)\left(-\ln Z -{\cal
E}\right) +\int d\vartheta~q(\vartheta)
E(\vartheta,\varphi)\nonumber\\
&=& -\ln Z -\int d\vartheta~q(\vartheta) {\cal E}(\vartheta)\nonumber \\&&+\int d\vartheta~q(\vartheta)E(\vartheta,\varphi)
\label{FE4-1}
\end{eqnarray}
where we have used the normalization condition, \eqnref{norm1} in the second step. The Gaussian integration involved in the first and second terms in equation~(\ref{FE4-1}) can be evaluated straightforwardly. Specifically, utilising \eqnref{PF}, the first term in
\eqnref{FE4-1} can be readily manipulated into 
\[ -\ln Z =  -\frac{1}{2} \left( \ln 2\pi\zeta \right).\]
Using \eqnref{PF} the  second term in \eqnref{FE4-1} becomes
\[ - \frac{1}{2\zeta}\int d\vartheta~q(\vartheta) ~(\vartheta-\mu)^2 \rightarrow -\frac{1}{2}. \]
The final term demands further technical consideration because the energy $E(\vartheta,\varphi)$ is still unspecified.  However, further simplifications can be made  by assuming that the R-density, \eqnref{total} is sharply peaked at its mean value $\mu$ (i.e., the Gaussian bell-shape is squeezed towards a delta function) and that ${E}(\vartheta,\varphi)$ is a smooth function of $\vartheta$. Under these assumptions we notice that the integration is appreciably non-zero only at the peaks. 
One can then use a Taylor expansion of ${E}(\vartheta,\varphi)$ around
$\vartheta=\mu$ with respect to a small increment, $\delta\vartheta$. 
Note:  while these assumptions permit a simple analytic model of the FEP, they have non-trivial implications for the interpretation of brain function so we return to this issue at the end of this section and in the Discussion.  This assumption brings about,
\begin{eqnarray*}
\label{taylor}
&&\int d\vartheta~ q(\vartheta){E}(\vartheta,\varphi), \\
&\approx & \int d\vartheta~ q(\vartheta)\left\{ {E}(\mu,\varphi) + 
\left[\frac{d {E}}{d\vartheta}\right]_\mu\delta\vartheta +
\frac{1}{2}\left[\frac{d^2{E}}{\partial
\vartheta^2}\right]_\mu \delta \vartheta^2\right\}.
\end{eqnarray*}
Now substituting back $\delta\vartheta = \vartheta-\mu$ we get,
\begin{eqnarray*}
&& \approx{E}(\mu,\varphi) + \left[\frac{\partial {E}} {\partial
\vartheta}\right]_\mu \int d\vartheta~ q(\vartheta)
(\vartheta-\mu)\\
&& + \frac{1}{2}\left[\frac{d^2{E}}
{d \vartheta^2}\right]_\mu \int
d\vartheta~ q(\vartheta)(\vartheta-\mu)^2.
\label{t2}
\end{eqnarray*}
Here the second term in the third line is zero identically because the integral equates to the mean. Furthermore recognising the expression for the variance in the  third term allows us to write
\begin{equation}
 \approx {E}(\mu,\varphi) + \frac{1}{2}\left[\frac{d^2{E}}
{d \vartheta^2}\right]_\mu\zeta.
\end{equation}
Where we identify $ {E}(\mu,\varphi) $ as the \textit{Laplace-encoded energy}. Substituting all terms derived so far into equation~(\ref{FE4-1})
furnishes an approximate expression for the IFE,
\begin{equation}\label{FE4-3}
F = {E}(\mu,\varphi) +
\frac{1}{2}\left(\left[\frac{d^2{E}}{d
\vartheta^2}\right]_\mu\zeta - \ln 2\pi\zeta -1 \right)
\end{equation}
which is now written as a \textit{function} (i.e., not a
functional) of the Gaussian means and variances, and sensory
inputs, i.e. $F = F(\mu,\zeta,\varphi)$. 
To simplify further we remove the dependence of the IFE on the variances by taking derivative of \eqnref{FE4-3} with respect $\zeta$ as follows:
\begin{eqnarray*}
d F &=& \frac{1}{2}\left\{\frac{d}{d
\zeta}
 \left(\left[\frac{d^2{E}}{d \vartheta^2}\right]_\mu\zeta\right)
 - \frac{1}{\zeta}\right\} d\zeta \\
 &=& \frac{1}{2}\left\{\left[\frac{d^2{E}}{d \vartheta_l}\right]_\mu
 - \frac{1}{\zeta}\right\} d\zeta.
\end{eqnarray*}
Minimising by demanding that  $d F\equiv 0$ one can get
\begin{equation}\label{gammastar}
\zeta^* = \left[\frac{d^2{E}}{d \vartheta^2}\right]_\mu^{-1}
\end{equation}
where the superscript in $\zeta^*$ indicates again that it is an optimal variance (i.e., it is the variance which optimizes the IFE). Substituting equation~(\ref{gammastar}) into equation~(\ref{FE4-3}) gives rise to the form of the IFE as
\begin{equation}\label{FE4-5}
F= {E}(\mu,\varphi) - \frac{1}{2}\ln \left\{2\pi
\zeta^*\right\} .
\end{equation}
The benefit of this process has been to recast the IFE in terms of a joint density $p(\mu, \varphi)$ over sensory data $\varphi$ and the R-density's sufficient statistics $\mu$, rather than a joint density over some (unspecified) environmental features $\vartheta$.  Note: this joint density amounts to an approximation of the G-density described in \eqnref{gdensity}; we shall examine the implementation of this density in detail in the next section.  Furthermore, under these assumptions the IFE only depends on Gaussian means (first-order Gaussian statistics) and sensory inputs, and not on variances (second-order  Gaussian statistics), which considerably simplifies the expression. It is possible to pursue an analogous derivation for the full multivariate Gaussian distribution under the more general assumption that the environment states only weakly covary, i.e., both the variance of, and covariances between, variables are small. Under this assumption the full R-density distribution is still tightly peaked and the Taylor expansion employed in \eqnref{taylor} is still valid.  

To get rid of the constant variance term in \eqnref{FE4-5}, we  write the Laplace-encoded energy for the full multivariate case, as an approximation for the full IFE as
\begin{equation}\label{minimum-energy}
{E}(\{\mu_\alpha\},\{\varphi_\alpha\}) =  -\ln p(\{\mu_\alpha\},\{\varphi_\alpha\}),
\end{equation}
where we define $\{\mu_\alpha\}$ and $\{\varphi_\alpha\}$ as vectors of brain states and sensory data respectively, corresponding to environmental variables $\{\vartheta_\alpha\}$  with $\alpha=1,2,\cdots,N$ indexing N variables. This equation for the Laplace-encoded energy serves as a general approximation for the IFE which we will use in the rest of this study.

Conceptually this expression suggests the brain represents only the most likely environmental causes of sensory data and not the details of their distribution per se .However, as we will see later, the brain also encodes uncertainties through (expectations about) variances  (inverse variances)  in the G-density.

Table 2 provides a glossary of mathematical objects involved in the Laplace encoding of the environmental states in the brain.

\begin{table} [b]
\begin{tabular}{p{1.7in}p{4.2in}}
\multicolumn{2}{l} {\bf Table 3. Mathematical glossary in the generative models}\\
\hline \hline \multicolumn{1}{l} {\bf Symbol} & \multicolumn{1}{l}
{\bf Name \& Description}\\ \hline

& \\
\underline{Simple model} & $p(\varphi,\mu)=p(\varphi|\mu)p(\mu)$ \\
& \\

$g(\mu;\theta)$ & Generative mapping between the brain states $\mu$ and the observed  data $\varphi$, paramterised by $\theta$ \\

$z$, $w$ & Random fluctuations represented by Gaussian noise \\

$\sigma_z$, $\sigma_w$ & That variance of these  fluctuations (the inverse of precisions) \\

$p(\varphi|\mu)$, $p(\mu)$ & Likelihood, prior of $\mu$, which together determine $p(\varphi,\mu)$ \\

& \\
\underline{Dynamical model} & $p(\varphi,\mu) = \prod_{n=0}^\infty
p(\varphi_{[n]}|\mu_{[n]})p(\mu_{[n+1]}|\mu_{[n]})$ \\
& \\

${\tilde \mu}$ & Brain states in generalized coordinates; an infinite vector
whose components are given by successive time-derivatives,
$\tilde\mu \equiv (\mu,\mu',\mu'',\cdots) \equiv
(\mu_{[0]},\mu_{[1]},\mu_{[2]},\cdots)$. \\

${\tilde \varphi}$ & Sensory data, similarly defined as $\tilde\varphi =
(\varphi,\varphi',\varphi'',\cdots)$. \\

$\varphi_{[n]} = g_{[n]}+z_{[n]}$ & Generalized mapping between the observed  data $\varphi$ and
the brain states $\mu$ at the dynamical order $n$ \\

$\mu_{[n+1]}= f_{[n]}+ w_{[n]}$ & Generalized equations of motion of
the brain state $\mu$ at the dynamical order $n$ \\

$g_{[n]}$, $f_{[n]}$ & Generative functions in the generalized coordinates \\

$p(\varphi_{[n]}|\mu_{[n]})$ & Likelihood of the generalized state $\mu_{[n]}$, given
the data $\varphi_{[n]}$ \\

$p(\mu_{[n+1]}|\mu_{[n]})$ & Gaussian prior of the generalized state $\mu_{[n]}$ \\

\hline

\end{tabular}
\end{table}

\section{The G-density: Encoding the brains beliefs about environmental causes}
\label{inference}
In the previous section we constructed an approximation of the IFE, which we called the Laplace-encoded energy, in terms of the approximate G-density $p(\mu,\varphi)$ where the environmental states $\vartheta$ have been replaced by the sufficient statistics $\mu$ of the R-density. In this section we consider how the brain could specify this G-density, and thus evaluate IFE. We start specifying a \textit{generative model} of the environmental causes of sensory data (informally, a description of causal dependencies in the environment and their relation to sensory signals). We then show how to move from these generative models to specification of the G-density, in terms of brain states and their expectations, and finally construct  expressions for the IFE. We develop various specifications of G-densities for both static and dynamic representations of the environment and derive the different expressions for IFE they imply. 

Table 3 provides a summary of the mathematical objects associated with the G-density in the simplest model and also its extension to the dynamical generative model.

\subsection{The simplest generative model}
\label{SingleLevel}
We first consider a simplified situation corresponding to an organism that believes  in an environment comprising of a single variable and a single sensory channel. To represent this environment the agent  utilise  a single brain state $\mu$ and sensory input $\varphi$. We then write the organisms belief about the environment directly in terms of a generative  mapping between  brain states  and  sensory data.  Note these equations will have a slightly strange construction because in reality sensory data is caused by environmental, not brain, states. However, writing the organism beliefs in this way will allow us to easily construct a generative density, see below. Specifically we assume the agent  believe its sensory is generated  as
\begin{equation}\label{map1}
\varphi = g(\mu;\theta) + z
\end{equation}
where $g$ is a linear or nonlinear function, parametrized by $\theta$ and $z$ is a random variable with zero mean  and variance $\sigma_z$.   Thus the organism believes its sensory  data is generated as non-linear mapping between environmental states (here denoted in terms of its  belief about environmental state $\mu$) with added noise. Similarly we specify the organism beliefs about how  environmental state are generated as 
\begin{equation}\label{brain-dynamics-form}
\mu = \bar\mu + w,
\end{equation}
where $\bar\mu$ is some fixed parameter and $w$ is random noise drawn from a Gaussian with zero mean and variance  $\sigma_w$. In other words, the organism takes the environment's future states to be history-independent, fluctuating around some mean value $\bar\mu$ which is given \textit{a priori} to the organism. There is a potential confusion here because equation \eqnref{brain-dynamics-form} describes a distribution over the brain state variable $\mu$, which itself represents the mean of some represented environmental state $\vartheta$.  Specifically it is worth reiterating that  $\bar\mu$  and  $\sigma_w$ are distinct from the sufficient statistics of the R-density $\mu$ and $\zeta$ [see \eqnref{Gaussian1}].  The variables $\bar\mu$ represent the organism's belief about the future state of the environment as  encoded in the G-density and $\sigma_w$ encodes the organism's confidence in its estimate of those future states. By contrast $\mu, \zeta$ belong to the R-density, encoding the organism's uncertain beliefs about its current environment $\vartheta$. As we will see in \secref{action}, there is conflict here because the organism's best estimate $\mu$ (the mean of its subjective distribution over $\vartheta$) may not be in line with its expectation $\bar\mu$ stemming from its model of environmental dynamics. 

To construct the generative density we assume that the noise $z$ is given as Gaussian, $[1/\sqrt{2\pi\sigma_z}] \exp \left\{ - z^2/(2\sigma_z)\right\}.$ Then, rewriting equation~(\ref{map1}) as
$z=\varphi-g(\mu;\theta)$, the functional form of the likelihood
$p(\varphi|\mu)$ can be written as
\begin{equation}\label{likelihood1}
p(\varphi|\mu) = \frac{1}{\sqrt{2\pi\sigma_z}} \exp \left\{ -
\left(\varphi - g(\mu;\theta)\right)^2/(2\sigma_z)\right\}.
\end{equation}
Assuming similar Gaussian noise for the random deviation
$w=\mu-\bar\mu$, in equation~(\ref{map2}), the prior density
$p(\mu)$ can be written as
\begin{equation}\label{prior1}
p(\mu) = \frac{1}{\sqrt{2\pi\sigma_w}} \exp \left\{ - \left(\mu -
\bar\mu\right)^2/(2\sigma_w)\right\}
\end{equation}
where  $\sigma_w$ is the variance.

Thus far, we have specified the likelihood and the prior of $\mu$ which together determine the G-density $p(\mu,\varphi)$ according to the identity,
\[ p(\mu,\varphi)=p(\varphi,\mu)=p(\varphi|\mu)p(\mu).\]
Next, we construct the Laplace-encoded energy  by substituting the likelihood and prior densities obtained above into \eqnref{minimum-energy} to get, up to a constant,
\begin{eqnarray}\label{value1}
{ E}(\mu,\varphi)
&=& - \ln p(\varphi|\mu) - \ln p(\mu)\\
&=& \frac{1}{2\sigma_z}\varepsilon_z^2 +
\frac{1}{2\sigma_w}\varepsilon_w^2 + \frac{1}{2}\ln\left(\sigma_z
\sigma_w\right),
\end{eqnarray}
where the auxiliary notations have been introduced as
\[ \varepsilon_z \equiv \varphi - g(\mu;\theta) \quad {\rm and}\quad
\varepsilon_w \equiv \mu - \bar\mu.
\]
which comprise a \textit{residual error} or a \textit{prediction error}
in the predictive coding terminology \cite{rao:1999}. The quantity $\varepsilon_z$ is a measure of the discrepancy between actual  $\varphi$ and the outcome of its prediction $ g(\mu;\theta)$. While $\varepsilon_w$ describes the extent to which $\mu$ itself deviates from its expectation $\bar\mu$. The former describes sensory prediction errors,  $\varepsilon_z$,  while the latter describe model predictions, $\varepsilon_w$ , (i.e., how brain states deviate from their expectation).  Each erro term is multiplied by the inverse of variance which weight the relative confidence of these term, i.e., how they contribute to the Laplace-encoded energy. We note in other works the inverse of variance, know as a precision, is used in these equations  perhaps to highlight that these terms weight the confidence, or preciseness, of the prediction.  However in this presentation we stick to more standard notation involving variances.    

The above calculation can be straightforwardly extended to the multivariate case. Specifically, we represent $\{\mu_{\alpha}\}$ as a row vector of $N$ brain states, and write their expectations as
\begin{equation*}
\mu_\alpha=  \bar{\mu}_{\alpha} + w_{\alpha}.
\end{equation*}
Here $\{{w_\alpha}\}$ is a row vector describing correlated noise sources, thus generally the fluctuations of each variable are not independent, which all have zero mean and covariance $\Sigma_{w}$.
We can the write a set of $N$ sensory inputs $\{\varphi_{\alpha}\}$ which depend on combination of these brain states in some nonlinear  way such that
\begin{equation}
\varphi_\alpha = g_\alpha(\mu_0, \mu_1,...,\mu_N) +z_\alpha.
\end{equation}
Again $\{{z_\alpha}\}$ are noise sources with zero mean and covariance $\Sigma_{z}$ and thus each sensory input may receive statistically correlated noise. Then, the prior over brain states may be represented as the multivariate correlated Gaussian density,
\begin{equation}
p(\{\mu_\alpha\}) = \frac{1}{\sqrt{(2\pi)^{N}|\Sigma_{w}|}}
\exp{ \left( -\frac{1}{2} \{\mu_\alpha - \bar{\mu}_{\alpha}\} \Sigma^{-1}_{w} \{\mu_\alpha - \bar{\mu}_{\alpha}\}^T
\right) },
\end{equation}
where $\{\mu_\alpha - \bar{\mu}_{\alpha}\}^T$ is the transpose of vector $\{\mu_\alpha - \bar{\mu}_{\alpha}\}$; $|\Sigma_w|$ and $\Sigma^{-1}_w$ are the determinant and the inverse of the covariance matrix $\Sigma_{w}$, respectively.
Similarly, we can write down the multivariate likelihood as
\begin{equation}
p(\{\varphi_\alpha\} |\{\mu_\alpha\}) =  \frac{1}{\sqrt{(2\pi)^{N}|\Sigma_z|}}
\exp{ \left( -\frac{1}{2} \{\varphi_\alpha- g_\alpha(\mu)\} \Sigma^{-1}_z \{\varphi_\alpha- g_\alpha(\mu)\}^T
\right) }.
\end{equation}
Now substituting these expressions into \eqnref{minimum-energy} we can get an expression of the Laplace-encoded energy as, up to an overall constant,
\begin{eqnarray}
E(\{\varphi_\alpha\},\{\mu_\alpha\})
&=& \frac{1}{2} \{\mu_\alpha - \bar{\mu}_{\alpha}\} \Sigma^{-1}_{w} \{\mu_\alpha - \bar{\mu}_{\alpha}\}^T + \frac{1}{2} \ln |\Sigma_{w}| \nonumber\\
&+& \frac{1}{2} \{\varphi_\alpha- g_\alpha(\mu)\} \Sigma^{-1}_z \{\varphi_\alpha- g_\alpha(\mu)\}^T + \frac{1}{2} \ln |\Sigma_z|. \label{gmulti1}
\end{eqnarray}
The above equation~(\ref{gmulti1}) contains non-trivial correlations among the brain variables and sensory data.
It is possible to pursue the full general case, e.g., see \cite{bogacz2015tutorial} for a nice tutorial on this, but we do not consider this here. Instead we can simplify on the assumption of statistical independence between environmental variables and between sensory inputs. Under this assumption the prior and likelihood are factorised into the simple forms, respectively,
\begin{eqnarray}
&& p(\{\mu_\alpha\}) = \prod_{\alpha=1}^N p(\mu_\alpha), \\
&& p(\{\varphi_\alpha\} |\{\mu_\alpha\}) = \prod_{\alpha=1}^N p(\varphi_\alpha |\{\mu_\alpha\}),
\end{eqnarray}
where probability densities are the uncorrelated Gaussians,
\begin{eqnarray*}
&& p(\{\mu_\alpha\}) = \prod_{\alpha=1}^{N}\frac{1}{\sqrt{2\pi\sigma_{z}^\alpha}} \exp \left\{ -
[\mu_\alpha-\bar{\mu}_\alpha ]^2/\left(2\sigma_{w}^\alpha\right)\right\}, \\
&& p(\{\varphi_\alpha\} | \{\mu_\alpha\}) = \prod_{\alpha=1}^{N} \frac{1}{\sqrt{2\pi\sigma_{w}^\alpha}} \exp \left\{ -
[\varphi_\alpha- g_\alpha(\mu) ]^2/\left(2\sigma_{z}^\alpha\right)\right\}.
\end{eqnarray*}
This gives the Laplace-encoded energy as
\begin{equation} \label{multi}
E(\{\varphi_\alpha\},\{\mu_\alpha\}) = \sum_{\alpha=1}^{N} \left[  \frac{({{\varepsilon}^\alpha_{w}})^2}{2\sigma_{w}^\alpha} + \frac{1}{2} \ln \sigma_{w}^\alpha \right] + \sum_{\alpha=1}^{N} \left[ \frac{({{\varepsilon}^\alpha_{z}})^2}{2\sigma_{z}^\alpha} + \frac{1}{2} \ln \sigma_{z}^\alpha \right],
\end{equation}
where the variances $\sigma_{w}^\alpha$ and $\sigma_{z}^\alpha$ are diagonal elements of the covariance matrices $\Sigma_w$ and $\Sigma_z$, respectively.
In \eqnref{multi} we have again used the auxiliary variables
\begin{eqnarray*}
{\varepsilon}_{w}^\alpha &=& \mu_{\alpha}-\bar{\mu}_{\alpha}, \\
{\varepsilon}_{z}^\alpha &=& \varphi_{\alpha}-g_{\alpha},
\end{eqnarray*}
The structure of \eqnref{multi} suggests that the Laplace-encoded energy, which is an approximation for the IFE, is a quadratic sum of the prediction-errors, modulated by the corresponding inverse variances, and an additional sum of the logarithm of the variances.                                                                      

\subsection{A dynamical generative model}
\label{DynamicalModel}
In the previous section we considered a simple generative model where an organism understood the environment to be effectively static. Here we extend the formulation to dynamic generative models which have the potential to support inference in dynamically changing environments. Again we start by examining a single sensory input $\varphi$ and a univariate brain state $\mu$. 
Here we assume that the agent's model of environmental dynamics (again expressed in terms of  brain states) follows not \eqnref{brain-dynamics-form}, but rather a Langevin-type equation \cite{Zwanzig:2001}
\begin{equation}\label{Langevin1}
\frac{d\mu}{dt} = f(\mu) + w
\end{equation}
where $f$ is a function of $\mu$ and $w$ is a random fluctuation. A dynamical generative model can then be obtained by combining the simple generative model, \eqnref{map1}, with \eqnref{Langevin1}.

The FEP utilizes the notions of \textit{generalized coordinates} and \textit{higher-order motion} \cite{Friston2008NeuroImage} to incorporate general forms of dynamics into the G-density. Generalised coordinates involve representing the state of a dynamical system in terms of increasingly higher order derivative of its state variables.  For example, generalized coordinates of a position variable may correspond to  bare `position' as well as its (unbounded) higher-order temporal derivatives (velocity, acceleration, jerk, and so on) allowing a more precise specification of a system's state \cite{Friston2008NeuroImage}.  To obtain these coordinates we simply take recursively higher order derivatives of both \eqnref{map1} and \eqnref{Langevin1}. 

For the sensory data:
\begin{eqnarray}\label{map2}
\varphi &=& g(\mu) + z \nonumber\\
\varphi' &=& \frac{\partial g}{\partial \mu}\mu' +z'\\
\varphi'' &=& \frac{\partial g}{\partial \mu}\mu''
+  z''\nonumber\\
&&\vdots\nonumber
\end{eqnarray}
where we have used the notations,
\[ \varphi' \equiv d\varphi/dt,\quad \mu'\equiv d\mu/dt,\quad \mu''\equiv
d^2\mu/dt^2, \quad etc. \] 
and where $z,z',...$ are the noises sources at each dynamic order. Here nonlinear derivative terms such as ${\mu'}^2$, $\mu'\mu''$, etc, have been neglected
under a \textit{local linearity assumption} \cite{Friston2006NeuroImage} and only linear terms have been collected. In some treatments of the FEP it is assumed that the noise sources are correlated \cite{Friston2008NeuroImage}. However, here, for the  clarity of the following derivations, we follow more standard state space models and assume each dynamical order receives independent noise, i.e, we assume the covariance between noise sources is zero.

Similarly, the Langevin equation, equation~(\ref{Langevin1}), is generalized as
\begin{eqnarray}\label{Langevin02}
\mu' &=& f(\mu) + w \nonumber\\
\mu'' &=& \frac{\partial f}{\partial \mu}\mu' + w'\\
\mu''' &=& \frac{\partial f}{\partial \mu}\mu''
+  w'' \nonumber\\
&&\vdots \nonumber
\end{eqnarray}
where again we have applied the local linearity approximation and we assume   each dynamical order receives independent noise denoted as $w,w',...$.
Here, it is convenient to denote the multi-dimensional sensory-data
$\tilde\varphi$ as
\[
\tilde\varphi =
(\varphi,\varphi',\varphi',\cdots)
\equiv(\varphi_{[0]},\varphi_{[1]},\varphi_{[2]},\cdots)
\]
and states $\tilde\mu$ as 
\begin{equation}\label{Gcoordinate}
\tilde\mu = (\mu,\mu',\mu'',\cdots) \equiv
(\mu_{[0]},\mu_{[1]},\mu_{[2]},\cdots),
\end{equation}
both being row vectors;
where the $n$th-components are defined to be
\[
{\varphi}_{[n]} \equiv \frac{d^{n}}{dt^{n}}\varphi = \varphi^\prime_{[n-1]} \quad {\rm and} \quad
{\mu}_{[n]} \equiv \frac{d^{n}}{dt^{n}}\mu = \mu^\prime_{[n-1]}.
\]
The generalized coordinates, equation~(\ref{Gcoordinate}), span the generalized state-space in mathematical terms.
In this state-space, a point represents an infinite-dimensional vector that encodes the instantaneous trajectory of a brain variable \cite{Friston2008One}.
By construction, the time-derivative of the state vector
$\tilde\mu$ becomes
\[
{\tilde\mu'}\equiv D\tilde \mu \equiv \frac{d}{dt}(\mu,\mu',\mu'',\cdots) =
(\mu',\mu'',\mu''' \cdots)
\equiv (\mu_{[1]},\mu_{[2]},\mu_{[3]},\cdots).
\]
The fluctuations in the generalized coordinates are written as
\[
\tilde z = (z,z', z'',\cdots) \equiv
(z_{[0]},z_{[1]},z_{[2]},\cdots),
\]
\[
\tilde w = (w,w',w'',\cdots) \equiv
(w_{[0]},w_{[1]},w_{[2]},\cdots).
\]
In addition, we denote the vectors associated with time-derivatives of the generative functions as
\[
\tilde g \equiv (g_{[0]},g_{[1]},g_{[2]},\cdots) \quad{\rm and}\quad
\tilde f \equiv (f_{[0]},f_{[1]},f_{[2]},\cdots)
\]
where the components are given as $g_{[0]} \equiv
g(\mu)$ and $f_{[0]}\equiv f(\mu)$, and for $n\ge 1$ as
\[
g_{[n]} \equiv \frac{\partial g}{\partial \mu}\mu_{[n]} \quad {\rm and}\quad
f_{[n]} \equiv \frac{\partial f}{\partial \mu}\mu_{[n]}.
\]
In terms of these constructs the infinite set of coupled equations~(\ref{map2})
and (\ref{Langevin02}) can be written in a compact form as
\begin{eqnarray}
\tilde\varphi &=& \tilde g + \tilde z \label{map3}\\
D{\tilde \mu} &=& \tilde f + \tilde w \label{Langevin2}
\end{eqnarray}
The generalized map, equation~(\ref{map3}), describes how the sensory data $\tilde\varphi$ are inferred by the representations of their causes $\tilde\mu$ at each dynamical order.  According to this map, the sensory data at a particular dynamical order $n$, i.e. $\varphi_{[n]}$, engages only with the same dynamical order of the brain states, i.e. $\mu_{[n]}$. The generalized equation of motion, equation~(\ref{Langevin2}), specifies the  coupling between  adjacent dynamical orders.

As before, in order to obtain the G-density we need to specify the likelihood of the sensory data $p(\tilde\varphi|\tilde\mu)$ and the prior $p(\tilde\mu)$. The statistical independence of  noise at each dynamical order means that we can write the likelihood as a product of conditional densities, i.e.,
\begin{eqnarray}\label{Dlikelihood1}
p(\tilde\varphi|\tilde\mu) &= & p(\varphi_{[0]},\varphi_{[1]},\varphi_{[2]},
\cdots|\mu_{[0]},\mu_{[1]},\mu_{[2]},\cdots)\nonumber\\
&=& \prod_{n=0}^\infty p(\varphi_{[n]}|\mu_{[n]}).
\end{eqnarray}
Assuming that the fluctuations at all dynamics orders, $z_{[n]}$, are induced by Gaussian noise, the conditional likelihood-density $p(\varphi^{[n]}|\mu^{[n]})$ is specified as
\[
p(\varphi_{[n]}|\mu_{[n]}) = \frac{1}{\sqrt{2\pi\sigma_{z[n]}}}
\exp \left[ - \left\{\varphi_{[n]}-g_{[n]}\right\}
^2/\left(2\sigma_{z[n]}\right)\right].
\]
Similarly, the postulate of the conditional independence of the generalized noises $w_{[n]}$ leads to a prior in the form
\begin{equation}\label{Dprior1}
p(\tilde\mu) = p(\mu_{[0]},\mu_{[1]},\mu_{[2]},\cdots) = \prod_{n=0}^\infty p(\mu_{[n+1]} | \mu_ {[n]})
\end{equation}
The form of the prior density at dynamical order $n$ is fixed by the assumption of Gaussian noise, which is then given as
\[
p(\mu_{[n+1]} | \mu_ {[n]}) = \frac{1}{\sqrt{2\pi\sigma_{w[n]}}} \exp \left[ -
\left\{\mu_{[n+1]}-f_{[n]} \right\}
^2/\left(2\sigma_{w[n]}\right)\right].
\]
Utilizing equations~(\ref{Dlikelihood1}) and (\ref{Dprior1}), the G-density is specified as
\begin{equation}\label{Dgenerative}
p(\tilde\varphi,\tilde\mu) = \prod_{n=0}^\infty p(\varphi_{[n]}|\mu_{[n]})p(\mu_{[n+1]} | \mu_ {[n]}).
\end{equation} 
Given the G-density, the Laplace-encoded energy can be calculated (equation~(\ref{minimum-energy})) to give, up to a constant,
\begin{eqnarray}\label{D-FFE1}
{E}(\tilde\mu,\tilde\varphi) &=& \sum_{n=0}^\infty\left\{
\frac{1}{2\sigma_{z[n]}}
[\varepsilon_{z[n]}]^2 + \frac{1}{2}\ln\sigma_{z[n]}\right\} \nonumber\\
&+& \sum_{n=0}^\infty\left\{\frac{1}{2\sigma_{w[n]}}
[\varepsilon_{w[n]}]^2 + \frac{1}{2}\ln\sigma_{w[n]}  \right\}
\end{eqnarray}
where $\varepsilon_{z[n]}$ and $\varepsilon_{w[n]}$ are $n$th component of the vectors $\tilde\varepsilon_z$ and $\tilde\varepsilon_w$, respectively, which have been defined to be
\[
\varepsilon_{z[n]}\equiv  \varphi_{[n]} - g_{[n]} \quad {\rm and} \quad
\varepsilon_{w[n]}\equiv \mu_{[n+1]} - f_{[n]}.
\]
As before, the auxiliary variables, $\varepsilon_{z[n]}$ and
$\varepsilon_{w[n]}$, encode \textit{prediction errors}:
$\varepsilon_{z[n]}$ is the error between the sensory data
$\varphi_{[n]}$ and its prediction $g_{[n]}$ at dynamical order
$n$. Likewise, $\varepsilon_{w[n]}$ measures the discrepancy
between the expected higher-order output $\mu_{[n+1]}$ and its generation $f_{[n]}$ from dynamical order $n$. 
Typically only dynamics up to finite order are considered. This can be done by setting the highest order term to random fluctuations, i.e.,
\[ \mu_{[n_{max}]} = w_{[n_{max}]}\]
where $w_{[n_{max}]}$ has  large variance; thus, the corresponding error term in \eqnref{D-FFE1} will be close to zero and effectively eliminated from the expression for the Laplace-encoded energy. 
In effect it means that the order below is unconstrained, and free to change in a way that best fits the incoming sensory data. This is related to the notion of empirical priors as discussed in \secref{HierarchicalModel}. Thus we have expressed 
Laplace-encoded energy for dynamics environment, which is an approximation for the IFE, is a quadratic sum of the sensory prediction-error,  $\varepsilon_{w[n]}$, and model prediction errors, $\varepsilon_{w[n]}$, across different dynamical orders. Again each error term is modulated by the corresponding variances describing the degree of certainty in those predictions.

Again its is straightforward we can generalise this to the multivariate case.
We set $\{\tilde\varphi_{\alpha}\} $ and $\{\tilde\mu_{\alpha}\} $ as vectors of brain states and rewrite \eqnrefs{map3} and (\ref{Langevin2}) as
\begin{eqnarray}
\tilde\varphi_{\alpha} &=& \tilde g_{\alpha} + \tilde z_{\alpha} \label{map3Mv2}\\
D{\tilde \mu}_{\alpha} &=& \tilde f_{\alpha} + \tilde w_{\alpha}, \label{Langevin2Mv2}
\end{eqnarray}
where $\alpha$ runs from $1$ to $N$.
Thus, \eqnref{D-FFE1} becomes
\begin{eqnarray}\label{D-FFE1-b}
{ E}(\{\tilde\mu_{\alpha}\} ,\{\tilde\varphi_{\alpha}\}) &=& \sum_{\alpha=1}^{N} \sum_{n=0}^\infty\left\{
\frac{1}{2\sigma_{z[n]}^{\alpha}}
[\varepsilon_{z[n]}^{\alpha}]^2 + \frac{1}{2}\ln\sigma_{z[n]}^{\alpha}\right\} \nonumber\\
&+& \sum_{\alpha=1}^{N} \sum_{n=0}^\infty\left\{\frac{1}{2\sigma_{w[n]}^{\alpha}}
[\varepsilon_{w[n]}^{\alpha}]^2 + \frac{1}{2}\ln\sigma_{w[n]}^{\alpha}  \right\}
\end{eqnarray}
where we have again used the auxiliary variables 
\begin{eqnarray}
\label{bfFC}
\varepsilon_{z[n]}^{\alpha} &\equiv&  \varphi_{\alpha[n]} - g_{\alpha[n]}\quad \\
\varepsilon_{w[n]}^{\alpha} &\equiv& \mu_{\alpha[n+1]}- f_{\alpha[n]}.
\end{eqnarray}
Thus this constitutes an  approximation of IFE for a multivariate system across arbitrary number of dynamical orders.

\section{IFE minimisation: How organisms infer environmental states}
\label{minimisation}
In the previous section we demonstrated how to go from a generative model, specifying the organism's  beliefs about the environment, to a generative density given expectations on brain states, and finally to an  expression for the IFE. In this section we discuss  how  organisms could  minimises IFE  to make the R-density a good approximation of the posterior and thus we begin to outline a full biologically plausible process theory.   In particular,  here,  we focus on how this minimisation could be implemented in neuronal dynamics of the brain outlining one particular \textit{process theory}.

Under the FEP it is proposed that the innate dynamics of the neural activity evolves in such a way as to minimise the IFE. Specifically, it is  suggested that brain states change in such  way that they implement a  gradient descent  scheme on IFE referred to as \textit{recognition dynamics}. 
Under the proposed gradient-descent scheme, a brain state $\mu_\alpha$ is updated between two sequential steps $t$ and $t+1$ as
\[
\mu_{\alpha}^{t+1} = \mu_{\alpha}^{t} -
\kappa\hat\mu_{\alpha}\cdot\nabla_{\mu_\alpha} { E}(\{\mu_\alpha\},\{\varphi_\alpha\})
\]
where $\kappa$ is the learning rate and $\hat\mu_{\alpha}$ is the unit vector along $\mu_{\alpha}$.  This process recursively modifies brain states in a way that follows  the gradient of Laplace-encoded energy. In the continuous limit the update $\mu^{\alpha}_{t+1} -
\mu^{\alpha}_{t}$ may be converted to a differential form as
\[
\mu_{\alpha}^{t+1} - \mu_{\alpha}^{t} \equiv \dot\mu_{\alpha}.
\]
Then, the above discrete updating-scheme can be transformed into a spatio-temporal differential equation,
\begin{equation}\label{Hgrad2-0}
\dot\mu_{\alpha} = -
\kappa\hat\mu_{\alpha}\cdot\nabla_{\mu_\alpha} { E}(\{\mu_\alpha\},\{\varphi_\alpha\}).
\end{equation}
The essence of the gradient descent method, as described in equation~(\ref{Hgrad2-0}), is that the minima of the objective function ${ E}$, i.e., the point where $\nabla_{\mu}{E}=0$, occur at the stationary solution when $\dot\mu_{\alpha}$ vanishes. Thus the dynamics of the brain states settle at a point where  the Laplace-encoded energy is minimized.

To update dynamical orders of the brain state $\mu_\alpha$, equation~(\ref{Hgrad2-0}) must be further generalized to give
\[
\mu_{\alpha[n+1]} - \mu_{\alpha[n]} \equiv -
\kappa\hat\mu_{\alpha[n]}\cdot\nabla_{\tilde \mu_\alpha} { E}
(\{{\tilde\mu}_\alpha\},\{{\tilde\varphi}_\alpha\})
\]
where $\hat\mu_{\alpha[n]}$ is the unit vector along $\mu_{\alpha[n]}$, $n$th-component of the generalized brain state $\tilde \mu_\alpha$ (\secref{DynamicalModel}).
Also, as before (equation~(\ref{Hgrad2-0})) the sequential dynamical order $(n,n+1)$ can be recast into a differential form to give
\begin{equation}\label{Dgradient-1}
\mu^\prime_{\alpha[n]} = - \kappa{\hat\mu}_{\alpha[n]}\cdot\nabla_{\tilde \mu_\alpha} {E}(\{{\tilde\mu}_\alpha\},\{{\tilde\varphi}_\alpha\}).
\end{equation}
Note that in order to be consistent with the definition of the generalised coordinates we have used the distinctive notation for the temporal derivative of dynamic update from the parametric update, equation~(\ref{Hgrad2-0}).
Here, we face a complication because the temporal derivative of the dynamical order $\mu_{\alpha[n]}$ is already contained within the generalized coordinates, \textit{i.e.}, $\mu^\prime_{\alpha[n]}= \mu_{\alpha[n+1]}$, in virtue
of the definition of the latter. 
Consequently, it is not possible to make $\mu^\prime_{\alpha[n]}$ vanish at any order, meaning that the gradient descent procedure is unable to construct a stationary solution at which the gradient of the Laplace-encoded energy vanishes. 
However, it can be argued that the motion of a point (velocity), i.e. $\dot {\tilde\mu}_\alpha$, in the generalized state-space is distinct from the `trajectory' encoded in the brain (flow velocity)
\cite{Friston2008NeuroImage,Friston2008NeuroImageTECH,Friston2008One}.
The latter object is denoted by $D{\tilde\mu_\alpha}$ where $D$ implies
also a time-derivative operator which, when acted on $\tilde\mu$, results in (see \secref{DynamicalModel})
\[
D{\tilde\mu}_\alpha \equiv (\mu_{\alpha[0]}^\prime,{\mu}_{\alpha[1]}^\prime,{\mu}_{\alpha[2]}^\prime \cdots)
\equiv (\mu_\alpha^\prime, \mu_\alpha^\second,\mu_\alpha^\third \cdots ),
\]
but is by this assumption distinct from the usual time-derivative $\dot{\tilde\mu}_\alpha$,
i.e. $\mu_{\alpha[n]}^\prime\neq \dot \mu_{\alpha[n]}$.
The term `velocity' here has been adapted by analogy with velocity in mechanics
in the sense that $\dot {\tilde\mu}_\alpha$ denotes first order time-derivative
of `position', namely the bare variable $\tilde\mu_\alpha$.
Prepared with this extra theoretical construct, the gradient descent scheme is restated in the FEP as
\begin{equation}\label{Dgrad2}
\dot{\mu}_{\alpha[n]} - D{\mu_{\alpha[n]}} = -
\kappa\hat\mu_{\alpha[n]}\cdot\nabla_{\tilde \mu_\alpha} {E}(\{{\tilde\mu}_\alpha\},\{{\tilde\varphi}_\alpha\})
\end{equation}
where $D{\mu_{\alpha[n]}}=\mu_{\alpha[n]}^\prime$.
According to this formulation, ${ E}$ is minimized with respect to the generalized state $\tilde\mu_\alpha$ when the `path of the mode' (generalized velocity) is equal to the `mode of the path' (average velocity), in other words the
gradient of ${E}$ vanishes when $\dot{\tilde\mu}_\alpha = D{\tilde\mu}_\alpha$. 
It is worth noting that in `static' situations where generalized motions are not required (see section~\ref{Parameters}), the concept of the `mode of the path' is not needed, i.e. $D{\tilde\mu}_\alpha\equiv 0$ by construction. 
In such situations we consider the relevant brain variables $\mu_\alpha$ to reach the desired minimum when there is no more temporal change in $\mu_\alpha$ in the usual sense, i.e. when $\dot\mu_\alpha = 0$.

In sum, these equations specify sets of first order ordinary differential equations that could be straightforwardly integrated by neuronal processing, e.g., they are very similar equations for firing rate dynamics in neural networks (e.g, see  \cite{haykin2004comprehensive}). Continuously integrating these equations in the presence of  stream of sensory data would make brain states continuously minimise  IFE and thus implement approximate inference on environmental states. Furthermore with additional assumption about there implementation \cite{friston2009predictive} they become  strongly analogous  to the predictive coding framework \cite{rao:1999}.

\section{Active inference}
\label{action}
A central appeal of the FEP is that it suggests not only an account of perceptual inference but also an account of action within the same framework: active inference. Specifically while perception minimises  IFE by changing brain states  to better predict sensory data, action instead acts on the environment to alter sensory input  to better fit sensory predictions. Thus action minimises IFE indirectly  by changing sensations.

In this section we describe a gradient-descent scheme analogous to that in the previous section but for action. To ground this idea for action, and combine it with the framework for perceptual inference discussed in previous sections, we  present an implementation of a simple agent-based model. 

Under the FEP action does not appear explicitly in the formulation of IFE but minimises IFE by changing sensory data. To evaluate this the brain must have a inverse model \cite{wolpert1997computational} of how sensory data change with action \cite{Friston2010BC}.  
Specifically, for a single brain state variable $\mu$ we write this as $\varphi=\varphi(a)$ where $a$ represents the action and $\varphi$ is a single sensory channel $\varphi$. 
Given this relationship we can then write the gradient of the Laplace-encoded energy with respect to action using the chain rule as,
\begin{equation}\label{chain}
  \frac{d{ E}({\mu},{\varphi})}{d a} \equiv  \frac{d\varphi}{da} \frac{\partial E({\mu},{\varphi})}{\partial\varphi}.
 \end{equation}
Thus we can write the same gradient decent scheme outlined in the last section to calculate the actions that minimise the Laplace-encoded energy as
\begin{equation}\label{grad-action0}
\dot a =
-\kappa_a\frac{d\varphi}{da} \frac{d{ E}({\mu},{\varphi})}{d\varphi}
\end{equation}
where $\kappa_a$ is the learning rate associated with action.

It is straightforward to write this gradient descent scheme for a vector of brain states in generalised coordinates as 
\begin{equation}\label{grad-action1}
\dot a =
-\kappa_a\sum_\alpha \frac{d\tilde\varphi_\alpha}{da} \cdot\nabla_{\tilde\varphi_\alpha}\ { E}(\{{\tilde\mu}_\alpha\},\{{\tilde\varphi}_\alpha\}).
\end{equation}

The idea that brains innately possess a inverse model, at first glance, seems somewhat troublesome. However, under the FEP the execution of motor control depends only on predictions about proprioceptors (internal sensors) which can be satisfied by  classic reflex arcs \cite{friston2011optimal,Friston2010BC}. On this reading  exteroceptive, and perhaps interoceptive \cite{seth2015cybernetic}, sensations are only indirectly minimised by action.  While a full assessment of this idea's implications is outside the remit of this work, it provides an interesting alternative to conventional notions of motor control, or behaviour optimisation, that rest on maximising or minimising value \cite{friston2011optimal}.

\begin{figure}
\begin{center}
\includegraphics[width=13cm]{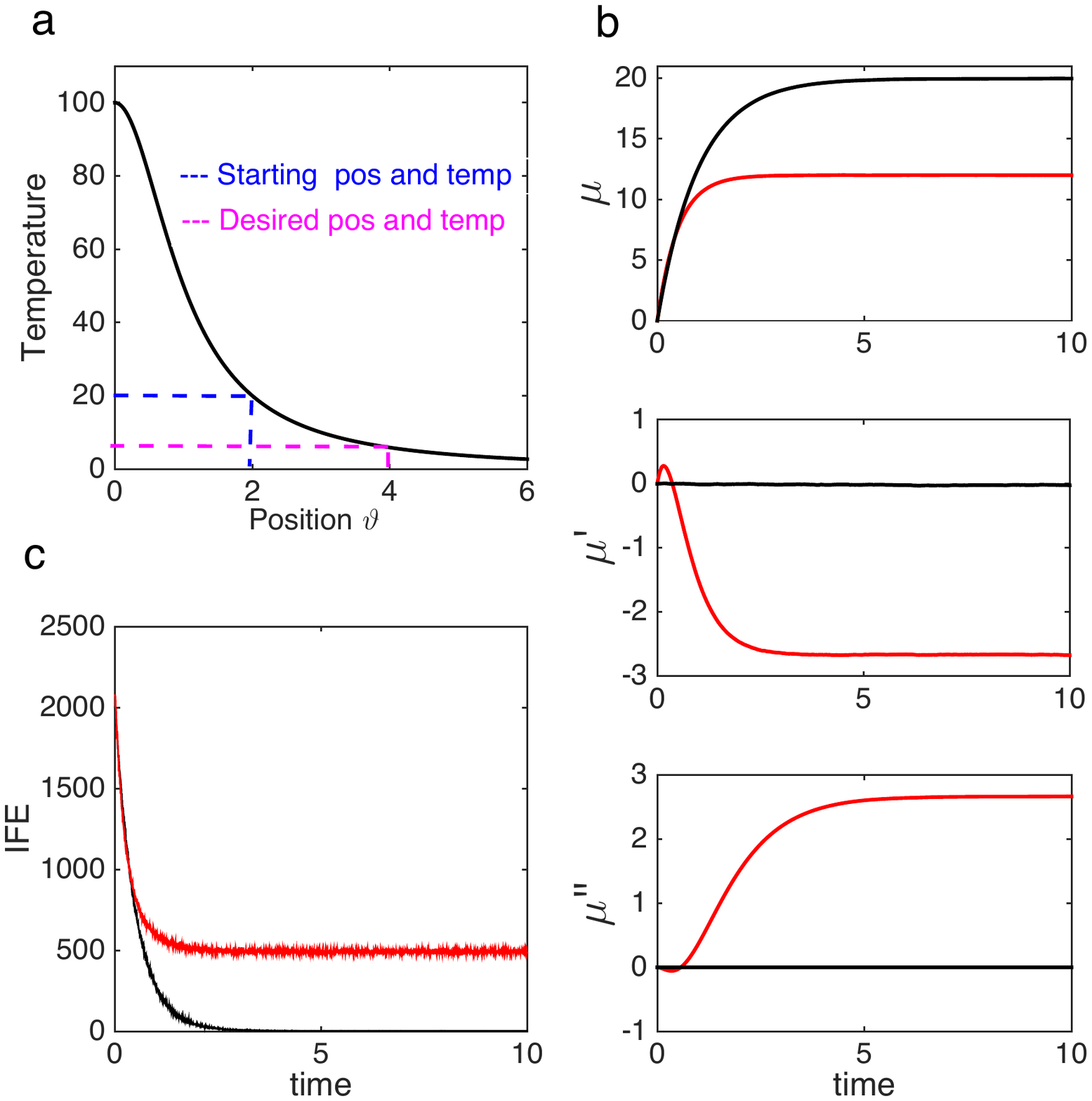}
\end{center}
\caption{Perceptual inference:
The agent's environment comprises a simple temperature gradient ({\bf{a}}), the blue and magenta lines give the actual and desired positions of the agent, respectively. The agent performs simple perceptual inference ({\bf{b}}), the dynamics of three generalized coordinates, $\mu$, $\mu'$ and $\mu''$, are given in the top, middle and bottom panels, respectively. Two conditions are shown, when the confidence in the sensory input is high (i.e. $\sigma_{z[i]}$ is small in comparison to  $\sigma_{w[i]}$), black line, and when confidence is equal between the internal model and sensory input, red line, respectively. IFE in both conditions monotonically decreases ({\bf{c}}): black and red traces, respectively. The tension between sensory input and internal model manifests a relatively high value of IFE ({\bf{c}}) (red curve), compared to the case where sensation has much higher confidence than the internal model (black curve).
}
\label{fig1}

\end{figure}

To give a concrete example of how perceptual and active inference work we present an implementation of a simple agent-based model. Specifically we present a model that comprises a mobile agent that must move to achieve some desired local temperature, $T_{desire}$.  The agent's environment, or \textit{generative process} \cite{Friston2010BC},  consists of a 1-dimensional plane and a simple temperature source. The agent's position on this plane is denoted by the environmental variable $\vartheta$ and the agent's temperature depends on its position in the following manner,
\begin{equation} \label{temp6}
T(\vartheta)= \frac{T_0}{\vartheta^2+1},
\end{equation}
where  $T_0$ is the temperature at the origin, i.e.,  this equation gives the true dynamics of the agents' environment (the environmental causes of its sensory signals).  The corresponding temperature gradient is readily given by,
\begin{equation*}
\frac{dT}{d\vartheta} = -T_0\frac{2\vartheta}{(\vartheta^2+1)^2} \equiv T_\vartheta.
\end{equation*}The temperature profile is depicted by the black line in \figref{fig1}a. We allow the agent to sense  both the local temperature and the temporal derivative of this temperature 
\begin{eqnarray}
\varphi &=& T +z_{gp} \\
\varphi^\prime &=& T_\vartheta \vartheta^\prime + z^\prime_{gp} \label{gprho}
\end{eqnarray}
where $z_{gp}$ and  $z^\prime_{gp}$ are normally distributed noise in the sensory readings.  
Note that the subscript $gp$ reminds us that this noise is a part of the agent's environment (rather than its brain model) described by the generative process. 

In this model the agent is presumed to sit on a flat frictionless plane and, thus, in the absence of action the agent is stationary. We allow the agent to set its own velocity by setting it equal to the action variable $a$ as,
\begin{equation}
\label{simpleAction}
\vartheta^\prime = a.
\end{equation}
The agent has brain state $\mu$ which represents the agents estimate of its temperature in the environment.  Following \eqnrefs{Langevin02}, we  write a generative model for the agent,  up to third order, as
\begin{eqnarray*}
\mu' &=& f(\mu) + w \quad {\rm where}\quad f(\mu)\equiv -\mu +T_{desire} \\
\mu'' &=& -\mu' +w'	\\
\mu''' &=& w'',
\end{eqnarray*}
where the third order term is just random fluctuations with large variance and thus is effectively eliminated from the expression for the Laplace-encoded energy, see \secref{DynamicalModel}.  
Following \eqnref{map2}, we write the agent's belief about it's sensory data only to first order as,
\begin{eqnarray*}
\varphi &=&g(\mu) +z \quad {\rm where}\quad g(\mu)\equiv \mu\\
\varphi' &=& \mu'+z' 
\end{eqnarray*}
Note the actual environment is not dynamic but the agent's belief about the environment is. Indeed, examining the agent's generative model we easily see that it possesses a stable equilibrium point at $T_{desire}$.  
In effect the agent believes in a environment where the forces it experiences naturally move it to its desired temperature, see \secref{overview} and  \cite{Friston2010BC}.  However,  these dynamics are different to those that describe the environment thus the agent must take action to make the environment conform.

We can write the Laplace-encoded energy, \eqnref{D-FFE1-b}, for this model, as
\begin{equation}\label{simLap}
E(\tilde\mu,\tilde\varphi) 
 =  \frac{1}{2}\left[ \frac{1}{\sigma_{z[0]}}(\varepsilon_{z[0]})^2 + \frac{1}{\sigma_{z[1]}}(\varepsilon_{z[1]})^2
+ \frac{1}{\sigma_{w[0]}}(\varepsilon_{w[0]})^2 + \frac{1}{\sigma_{w[1]}}(\varepsilon_{w[1]})^2 \right],
\end{equation}
where the various error terms are given as
\begin{eqnarray*}
\varepsilon_{z[0]} &=& \varphi - \mu \\
\varepsilon_{z[1]} &=& {\varphi}'-{\mu}' \\
\varepsilon_{w[0]} &=& \mu' +\mu-T_{desire} \\
\varepsilon_{w[1]} &=& \mu'' + \mu'.
\end{eqnarray*}
Also, $\sigma_{z[0]}$, $\sigma_{z[1]}$, $\sigma_{w[0]}$, and $\sigma_{w[1]}$ in \eqnref{simLap} are the variances corresponding to the noise terms $z$, $z'$, $w$, and $w'$, respectively.
In addition we have dropped logarithm of variance terms, see \eqnref{value1}  because they play no role when we minimise these equations with respect to  the brain variable $\mu$.

Note, that the noise terms in the agents internal model are distinct from those in \eqnref{gprho} and represent the agents beliefs about the noise on environmental states and sensory data rather than the actual noise on these variables. As we will see these terms effectively represent the confidence of the  agent in its own sensory input.  

Using the gradient decent scheme described in \eqnref{Dgrad2} we write the recognition dynamics as

\begin{eqnarray} \label{braindyn}
\dot\mu &=& \mu' - \kappa_a \left [ -\frac{\varepsilon_{z[0]}}{\sigma_{z[0]}}  +  \frac{\varepsilon_{w[0]}}{\sigma_{w[0]}} \right]\nonumber \\
\dot{\mu}' &=& \mu'' - \kappa_a \left [ -\frac{\varepsilon_{z[1]}}{\sigma_{z[1]}}  + \frac{\varepsilon_{w[0]}}{\sigma_{w[0]}} + \frac{\varepsilon_{w[1]}}{\sigma_{w[1]}}  \right ] \\
\dot{\mu}'' &=& - \kappa_a \frac{\varepsilon_{w[1]}}{\sigma_{w[1]}}. \nonumber
\end{eqnarray}
Here we have considered generalised coordinates up to second order only.  To allow the agent to perform action we must provide it with an inverse model, which we assume is hard-wired \cite{Friston2010BC}. Replacing the agent's velocity with the action variable $a$ in \eqnref{gprho} we specify this as
\begin{equation}
\frac{d {\varphi'}}{d a} = \frac{d}{da} \left( a T_\vartheta + z_{gp}^\prime \right) = T_\vartheta.
\end{equation}
Effectively the agent believes that action changes the  temperature in a way that is consistent with it's beliefs about the  temperature gradient. Given this inverse model we can write down the minimisation scheme for action as.
\begin{equation}
\label{actiondyn}
\dot{a} =  -\kappa_a \left[ \frac{d\varphi}{da} \frac{\partial E}{\partial \varphi} + \frac{d\varphi'}{da}\frac{\partial E}{\partial \varphi'} \right] = -\kappa_a T_\vartheta\frac{\varepsilon_{z[1]}}{\sigma_{z[1]}}.
\end{equation}
Thus, \eqnrefs{braindyn} through (\ref{actiondyn}) describe the complete agent-environment system and can be straightforwardly integrated (see \ref{code} for details).  

\begin{figure}
\begin{center}
\includegraphics[width=13cm]{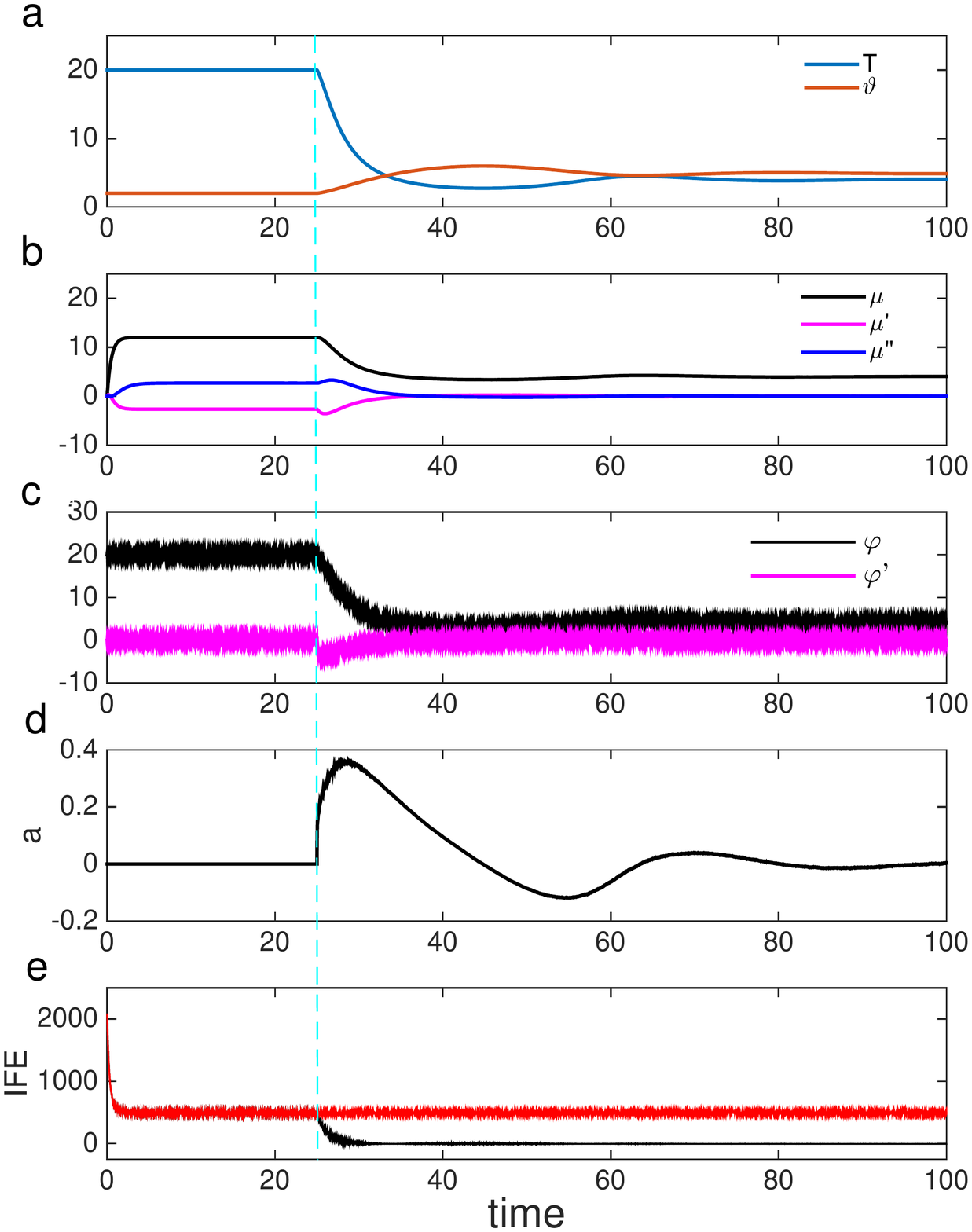}
\caption{Perceptual and active inference: An agent with equal confidence in its internal model and sensory input $\sigma_{z[i]} =\sigma_{w[i]}=1$ is allowed to act at $t=25$.  The agent acts, see ({\bf{d}}), to alter its position, see ({\bf{a}}: orange line), to bring down its initial temperature ($T=20$) to the desired temperature ($T=T_{desire}=4$), see ({\bf{a}}: blue line). It does this by  bringing its sensory data ({\bf{c}}) in line with its desire, i.e., $\varphi= T_{desire}$ and thus the brain state becomes equal to its desired state, see ({\bf{b}}). IFE was calculated in the presence and absence of the onset of action at $t=25$, see {\bf{e}}, black and red lines, respectively. First IFE is reduced by inference ($t<25$), then later through action ({\bf{e}}: black line).}
\label{fig2}
\end{center}
\end{figure}

\figref{fig1} shows the behaviour of the agent in the absence of action, i.e., when the agent is unable to move. We examine two conditions. In a first condition the agent's sensory variances $\sigma_{z[0]}$, $\sigma_{z[1]}$ are several orders of magnitude smaller than model variances $\sigma_{w[0]}$ and $\sigma_{w[1]}$. Thus the agent has higher confidence (see \secref{SingleLevel}) in sensory input than in its internal model.  Under this condition the agent successfully infers both the local temperature  and its corresponding derivatives, see \figref{fig1}b black lines. In effect the agent ignores its internal model and the gradient decent scheme is equivalent to a least mean square estimation on the sensory data, see supplied code in \ref{code}. In a second condition, see  \figref{fig2} red lines, we equally balance internal model and sensory variances ($\sigma_{z[i]}= \sigma_{w[i]},\ i=0,1$). Now minimisation of IFE cannot satisfy both sensory perception and predictions of the agent's internal model, i.e., what the agent perceives is in conflict with what it desires. Thus the inferred local temperature sits somewhere between its desired and sensed temperature, see \figref{fig1}b. 

In \figref{fig2}, after an initial period, the agent is allowed to act according to \eqnref{actiondyn}. It does so by changing the environment to bring it in line with with sensory predictions and the desires encoded within its dynamic model, i.e., the agent moves toward the desired temperatures.   

The reduction of surprisal can be  quantified as the difference between  the Laplace-encoded energy (and thus IFE) in presence and absence of action, i.e., the difference between  black and red traces in \figref{fig2}e, respectively.  Specifically, it is the portion of the IFE that must be minimised by  acting on the environment rather than through optimisation of the  agent's environment model. We leave a more explicit quantification of the dynamics of surprisal for future work.

In summary we have presented an example of an agent performing a very simple task under the FEP. The model demonstrates how the minimisation of IFE can underpin both perception and action. Furthermore, it shows how a tension between desires and perception can be reconciled through action. Many other agent based implementations of the FEP have been presented in the literature, .e.g. \cite{Friston2010BC}, which can be constructed in a similarly simplistic way.

\section{Hierarchical Inference and Learning}
\label{learning}

In the previous sections we developed the FEP for organisms given simple dynamical generative models. We then investigated the emergence of behaviour in a simulated organism (agent) furnished with an appropriate generative model of a simple environment. The assumption here was that organisms possess some knowledge or beliefs of about how the environment works a priori, in the form of a pre-specified generative model. However, another promise of the FEP is the ability to learn and infer arbitrary environmental dynamics \cite{Friston2008One}.  To achieve this it is suggested that brain starts out with a very general hierarchical generative model of environmental dynamics which is moulded and refined through experience.  The advantage of using hierarchical models, as we will see, is that they suggest a way of avoiding specifying an explicit and fixed prior, and thus can implement empirical Bayes \cite{Casella:2002}. In this section we first provide a description of a  hierarchical G-density which is capable of empirical Bayes \cite{Casella:2002}. We then combine this with dynamical generative model described in \eqnref{D-FFE1-b} to define what we shall call the \textit{full construct}. We  go on to describe how appropriate  parameters and hyperparameters of the G-density for given world could be discovered through learning. We finish this section by showing how action could be described in this construct. 
\subsection{Hierarchical generative model}
\label{HierarchicalModel}

\begin{table}[b]
\begin{tabular}{p{1.7in}p{4.2in}}
\multicolumn{2}{l} {\bf Table 4. Mathematical constructs in the hierarchical generative model}\\
\hline \hline \multicolumn{1}{l} {\bf Symbol} & \multicolumn{1}{l}
{\bf Name \& Description}\\ \hline
& \\
\underline{Hierarchical model} & $p(\varphi,\mu)= p(\mu^{(M)}) \prod_{i=0}^{M-1} p(\mu^{(i)}|\mu^{(i+1)}) $  \\
& \\
$\mu^{(i)}$ & Brain states at cortical layer $i$ ($i=1,2,\cdots, M$); $\mu^{(0)}\equiv \varphi$ denotes
the sensory data which reside at the lowest cortical layer. \\

$g^{(i)}(\mu^{(i)})$ & Generative map (or function) of the brain state $\mu^{(i)}$ to estimate
one-level lower state $\mu^{(i-1)}$ in the cortical hierarchy via
$\mu^{(i-1)} = g^{(i)}(\mu^{(i)}) + z^{(i-1)}$; where $z^{(i-1)}$ is Gaussian noise. \\

$p(\mu^{(i)}|\mu^{(i+1)})$ & Likelihood of $\mu^{(i)}$ given a value for $\mu^{(i+1)}$;  which acts as a prior
for $p(\mu^{(i-1)}|\mu^{(i)})$ in the cortical hierarchy. \\

$p(\mu^{(M)})$ & Probabilistic representation of brain states
at the highest layer, which forms the highest prior. \\
\hline

\end{tabular}
\end{table}

A key challenge for Bayesian inference models is how to specify the priors. Hierarchical models provide a powerful response to this challenge, in which higher levels can provide empirical priors or constraints on lower levels \cite{kass:1989}.  In the FEP, hierarchical models are mapped onto the hierarchical organisation of the cortex \cite{Zeki-Shipp1988,Felleman-Essen1991}, which requires extension of the simple generative model described above.

We denote $\mu^{(i)}$ as a brain state at hierarchical level $i$ and we assume $M$ cortical levels, with $i=1$ the \textit{lowest} level
and $i=M$ as the \textit{highest}. Then, the hierarchical model may be written explicitly as \cite{Friston2008One}
\begin{eqnarray*}
\varphi &=& g^{(1)}(\mu^{(1)}) + z^{(0)}\\
\mu^{(1)} &=& g^{(2)}(\mu^{(2)}) + z^{(1)}\\
\mu^{(2)} &=& \cdots \\
&&\vdots\\
\mu^{({M)}} &=& z^{(M)}
\end{eqnarray*}
which can be written compactly as
\begin{equation}\label{hierarchy1}
\mu^{(i)} = g^{(i+1)}(\mu^{(i+1)}) + z^{(i)}
\end{equation}
where $i$ runs through $1,2,\cdots M$. We further assume that the
sensory data $\varphi$ reside exclusively at the lowest cortical
level $\mu^{(1)}$ and dynamics at the highest level $\mu^{(M)}$ are
governed by a random fluctuation $z^{(M)}$, i.e:
\begin{equation}\label{aks1}
\mu^{(0)} \equiv \varphi \quad{\rm and} \quad g^{({M}+1)}\equiv 0.
\end{equation}
The hierarchy equation~(\ref{hierarchy1}) specifies that a cortical
state $\mu^{(i)}$ is connected to higher level $\mu^{(i+1)}$ through
the generative function $g^{(i+1)}$.  The fluctuations $z^{(i)}$ exist at each level, in particular
$z^{(0)}$ designating the observation noise at the sensory interface, and are
assumed to be statistically independent.

Having defined the hierarchical model, one can write the
corresponding G-density as
\begin{eqnarray}\label{generative2}
p(\varphi,\mu)&=& p(\mu^{(0)}|\mu^{(1)},\mu^{(2)},\cdots,\mu^{(N)})p(\mu^{(1)},
\mu^{(2)},\cdots,\mu^{(M)})\nonumber\\
&\equiv& p(\mu^{(0)}|\mu^{(1)})p(\mu^{(1)}|\mu^{(2)})\cdots p(\mu^{({M}-1)}|\mu^{(M)})
p(\mu^{(M)}).
\end{eqnarray}
The second step in equation~(\ref{generative2}) assumes that the
transition probabilities from higher levels to lower levels are
Markovian. Consequently, equation~(\ref{generative2}) asserts that
the likelihood of a level, for instance $p(\mu^{(i)}|\mu^{(i+1)})$,
serves as a prior density for the level immediately below, $i-1$.
The prior at the highest level $p(\mu^{(M)})$ contains information
only with respect to its spontaneous noise, which may be given by a
Gaussian form
\begin{equation}\label{prior2}
p(\mu^{(M)}) = \frac{1}{\sqrt{2\pi\sigma_z^{(M)}}} \exp \left\{ -
[\mu^{(M)}]^2/\left(2\sigma_z^{(M)}\right)\right\}
\end{equation}
where the mean has been assumed to be zero and $\sigma_z^{(M)}$ is
the variance.  We shall further assume that the Gaussian noises are
responsible for the (statistically independent) fluctuations at
all hierarchical levels. Accordingly, the likelihoods
$p(\mu^{(i)}|\mu^{(i+1)})$ are given as
\begin{equation}\label{likelihood2}
p(\mu^{(i)}|\mu^{(i+1)}) = \frac{1}{\sqrt{2\pi\sigma_z^{(i)}}}
\exp \left[ - \left\{\mu^{(i)} -
g^{(i+1)}(\mu^{(i+1)})\right\}^2/\left\{2\sigma_z^{(i)}
\right\}\right].
\end{equation}
and the G-density reduces to
\begin{equation}\label{generative3}
p(\varphi,\mu) = \left[\prod_{i=0}^{M}
\frac{1}{\sqrt{2\pi\sigma_z^{(i)}}}\right] \exp\left(
-\sum_{i=0}^{M} \frac{1}{2\sigma_z^{(i)}} [\varepsilon^{(i+1)}]^2
\right)
\end{equation}
where the auxiliary variables $\varepsilon^{(i)}$ have been introduced as
\begin{equation}\label{prediction2}
\varepsilon^{(i)} \equiv \mu^{(i-1)} - g^{(i)}(\mu^{(i)}).
\end{equation}
The quantity $\varepsilon^{(i)}$ measures  the discrepancy between
the prediction (estimation) at a given level $\mu^{(i)}$ via $g^{(i)}$
and $\mu^{(i-1)}$ at a lower-level, which comprises a \textit{prediction error}.

Finally, by substituting the G-density, constructed in equation~(\ref{generative3}), into
equation~(\ref{minimum-energy}), after a simple manipulation,
the Laplace-encoded energy ${ E}$ is given up to a constant as
\begin{equation}\label{FFE1}
{ E}(\mu,\varphi) = \sum_{i=0}^{M}\left\{
\frac{1}{2\sigma_z^{(i)}} [\varepsilon^{(i+1)}]^2 +
\frac{1}{2}\ln\sigma_z^{(i)}   \right\}.
\end{equation}
The variance of the noise at the top level of hierarchy is typically assumed to be large and thus  the corresponding term in the Laplace-encoded energy \eqnref{FFE1} is approximately zero. As with the higher dynamical orders discussed above \secref{DynamicalModel} this means that the level below is effectively unconstrained (has no prior) and thus this type of inference constitutes an example of empirical Bayes \cite{Casella:2002}.

Table 4 itemizes the mathematical objects associated with the hierarchical generative model. 

\subsection{Combining hierarchical and dynamical models:
 The full construct}
\label{Full-construct}

\begin{table}[b]
\begin{tabular}{p{1.7in}p{4.2in}}
\multicolumn{2}{l} {\bf Table 5. Mathematical constructs in the full generative model}\\
\hline \hline \multicolumn{1}{l} {\bf Symbol} & \multicolumn{1}{l}
{\bf Name \& Description}\\ \hline
& \\
\underline{Full construct} &   $p({\tilde\varphi}_\alpha,{\tilde\mu}_\alpha)=p({\tilde x}_\alpha^{(M)},{\tilde v}_\alpha^{(M)})\prod_{i=0}^{{M}-1}
p({\tilde x}_\alpha^{(i)}|{\tilde v}_\alpha^{(i)})p({\tilde v}_\alpha^{(i)}|{\tilde x}_\alpha^{(i+1)}, {\tilde v}_\alpha^{(i+1)})$ \\
& \\

$\tilde \mu_\alpha^{(i)}$ & Brain state $\alpha$ in cortical layer $i$ in generalized
coordinates, whose $n$th component is denoted as $\mu^{(i)}_{\alpha[n]}$.\\

${\tilde x}_\alpha^{(i)}$, ${\tilde v}_\alpha^{(i)}$ & Two distinctive neuronal representations,
${\tilde\mu}^{(i)}_\alpha = ({\tilde x}_\alpha^{(i)}, {\tilde v}_\alpha^{(i)})$;
designated as hidden and causal states, respectively. \\
${\tilde g_\alpha^{(i)}}$ & Generative map of the causal state ${\tilde v}_\alpha^{(i)}$ to learn the state
one layer below, ${\tilde v}_\alpha^{(i-1)} = {\tilde g}_\alpha^{(i)} ( {\tilde x}_\alpha^{(i)},{\tilde v}_\alpha^{(i)} )
+ {\tilde z}_\alpha^{(i-1)}$. \\

${\tilde f}_\alpha^{(i)}$ & Generative function which induces the Langevin-type equation of motion of the hidden state ${\tilde x}_\alpha^{(i)}$, $\dot{\tilde x}_\alpha^{(i)} = {\tilde f}_\alpha^{(i)}( {\tilde x}_\alpha^{(i)},{\tilde v}_\alpha^{(i)} )+ {\tilde w}_\alpha^{(i)}$. \\

${\tilde z}_\alpha^{(i)}$, ${\tilde w}_\alpha^{(i)}$ & Random fluctuations treated as Gaussian noise. \\

$p({\tilde x}_\alpha^{(M)}, {\tilde v}_\alpha^{(M)})$ & Prior density of the brain state $\tilde\mu_\alpha$ at the highest cortical layer $(M)$. \\

$p({\tilde x}_\alpha^{(i)}|{\tilde v}_\alpha^{(i)})$ & Probabilistic representation of the intra-layer dynamics of hidden states ${\tilde x}_\alpha^{(i)}$ conditioned on the causal state ${\tilde v}_\alpha^{(i)}$ via ${\tilde f}_\alpha^{(i)}$; dynamic transition from order $n$ to $n+1$ is hypothesized as the Gaussian fluctuation of  $w^{(i)}_{\alpha[n]} = x^{(i)}_{\alpha[n+1]} - f^{(i)}_{\alpha[n]}$. \\

$p({\tilde v}_\alpha^{(i)}|{\tilde x}_\alpha^{(i+1)}, {\tilde v}_\alpha^{(i+1)})$ & Likelihood density of the causal state ${\tilde v}^{(i)}$ which serves as a prior for one layer lower density, representing statistically the inter-layer map between two successive causal states, $z^{(i)}_{\alpha[n]} = v^{(i)}_{\alpha[n]}-g^{(i+1)}_{\alpha[n]}$, by the Gaussian fluctuation. \\

\hline
\end{tabular}
\end{table}
We now combine the dynamical structure and the multivariate brain states in a single expression.  First we note that under the FEP brain states representing neuronal activity $\mu_\alpha$ are divided into the \textit{hidden} states $x_\alpha$ and the \textit{causal} states $v_\alpha$, 
\[ \mu_\alpha= (x_\alpha, v_\alpha).\]
Then, the full FEP implementation can be derived formally by extending equations~(\ref{map3Mv2}) and (\ref{Langevin2Mv2}) (\eqnref{hierarchy1})
 \begin{eqnarray}
{\tilde v}^{(i)}_\alpha &=& {\tilde g}^{(i+1)}_\alpha ( {\tilde x}^{(i+1)}_\alpha,{\tilde v}^{(i+1)}_\alpha )
+ {\tilde z}^{(i)}_\alpha,\quad i=0,1,\cdots.M \label{inter-layer2}\\
D{\tilde x}^{(i)}_\alpha &=& {\tilde f}^{(i)}_\alpha( {\tilde x}^{(i)}_\alpha,{\tilde v}^{(i)}_\alpha )
+ {\tilde w}^{(i)}_\alpha, \quad i=1,2.\cdots,M \label{intra-layer2}
\end{eqnarray}
where the brain-state index runs through $\alpha=1,2,\cdots, N$ and ${\tilde v}_\alpha^{(0)}$ designates the sensory data at the lowest cortical layer, $i=1$.
Inter-layer hierarchical links are made through the causal states and intra-hierarchical layer dynamics through the hidden states. 
The generalized coordinates of neuronal brain state $\alpha$ in hierarchical layer $i$ are given by the
infinite-dimensional vectors
\[
\tilde x^{(i)}_\alpha \equiv
(x^{(i)}_{\alpha[0]},x^{(i)}_{\alpha[1]},x^{(i)}_{\alpha[2]},\cdots) \quad{\rm and}
\quad \tilde v^{(i)}_\alpha \equiv (v^{(i)}_{\alpha[0]},v^{(i)}_{\alpha[1]},v^{(i)}_{\alpha[2]},\cdots)
\]
where the components are labelled by the subscripts $[n]$,
$n=0,1,\cdots,\infty$.
Note that we have introduced different notations in the vector components:  The subscript $\alpha$ for brain states at a given hierarchical level, the superscript $(i)$ for the hierarchical indices, and the subscript $[n]$ for the dynamical orders.
Recall that the $n$-th component of the vector ${\tilde x}^{(i)}_\alpha$ and ${\tilde v}^{(i)}_\alpha$ are time-derivatives of order $n$, namely
\[
x^{(i)}_{\alpha[n]} \equiv \frac{d^{n}}{dt^{n}}x^{(i)}_\alpha \quad{\rm
and} \quad  v^{(i)}_{\alpha[n]} \equiv \frac{d^{n}}{dt^{n}}v^{(i)}_\alpha.
\]
The other mathematical quantities in equations~(\ref{inter-layer2}) and (\ref{intra-layer2}) are given
explicitly as:
\[
D{\tilde x}^{(i)}_\alpha =
(x^{(i)}_{\alpha[1]},x^{(i)}_{\alpha[2]},x^{(i)}_{\alpha[3]},\cdots),
\]
\[
{\tilde z}^{(i)}_\alpha \equiv (z^{(i)}_{\alpha[0]},z^{(i)}_{\alpha[1]},z^{(i)}_{\alpha[2]},\cdots),\quad {\rm
and}\quad {\tilde w}^{(i)}_\alpha \equiv
(w^{(i)}_{\alpha[0]},w^{(i)}_{\alpha[1]},w^{(i)}_{\alpha[2]},\cdots).
\]
The generative functions appearing in equations~(\ref{inter-layer2}) and (\ref{intra-layer2}) are
specified for $n\ge 1$, under the local-linearity assumption, as
\[
g_{\alpha[n]}({x}^{(i+1)}_{\alpha[n]},{v}^{(i+1)}_{\alpha[n]}) \equiv
\frac{\partial g} {\partial v^{(i+1)}_{\alpha[n]}} {v}_{\alpha[n]}^{(i+1)} 
\equiv {g}_{\alpha[n]}^{(i+1)}
\]
and
\[
{f}_{\alpha[n]}({x}^{(i)}_{\alpha[n]},{v}^{(i)}_{\alpha[n]})  \equiv
\frac{\partial f} {\partial x^{(i)}_{\alpha[n]}} {x}_{\alpha[n]}^{(i)} \equiv
{f}_{\alpha[n]}^{(i)}.
\]
For the lowest dynamical order of $n=0$,
\[
{g}_{\alpha[0]}^{(i+1)}=g( {x}^{(i+1)}_{\alpha[0]},{v}^{(i+1)}_{\alpha[0]}) \quad{\rm
and}\quad {f}_{\alpha[0]}^{(i)}=f({x}^{(i)}_{\alpha[0]},{v}^{(i)}_{\alpha[0]}).
\]

It is evident from equation~(\ref{inter-layer2}) that the causal states ${\tilde v}^{(i)}_\alpha$ at one hierarchical layer are predicted from states at one level higher in the hierarchy ${\tilde
v}^{(i+1)}_\alpha$ via the map ${\tilde g}^{(i+1)}_\alpha$: ${\tilde z}^{(i)}_\alpha$ specifies the fluctuations associated with these inter-layer links.
Equation~(\ref{intra-layer2}) asserts that the dynamical transitions of the hidden states ${\tilde x}^{(i)}_\alpha$ are induced \textit{within} a given hierarchical layer via ${\tilde f}^{(i)}_\alpha$:
The corresponding fluctuations are given by ${\tilde w}^{(i)}_\alpha$.
In order to describe these transitions more transparently, we spell out equations~(\ref{inter-layer2}) and
(\ref{intra-layer2}) explicitly:
\begin{eqnarray*}
\tilde v^{(0)}_\alpha = {\tilde g}^{(1)}_\alpha({\tilde x}^{(1)}_\alpha,{\tilde v}^{(1)}_\alpha) +
{\tilde z}^{(0)}_\alpha &\quad& \dot{\tilde x}^{(1)}_\alpha = {\tilde f}^{(1)}_\alpha
({\tilde x}^{(1)}_\alpha,{\tilde v}^{(1)}_\alpha) + {\tilde w}^{(1)}_\alpha \\
\tilde v^{(1)}_\alpha  = {\tilde g}^{(2)}_\alpha({\tilde x}^{(2)}_\alpha, {\tilde v}^{(2)}_\alpha) +
{\tilde z}^{(1)}_\alpha &\quad& \dot{\tilde x}^{(2)}_\alpha = {\tilde f}^{(2)}_\alpha
({\tilde x}^{(2)}_\alpha,{\tilde v}^{(2)}_\alpha) + {\tilde w}^{(2)}_\alpha \\
\qquad\vdots &\quad& \qquad \quad\vdots\\
\tilde v^{({M}-1)}_\alpha  = {\tilde g}^{({M})}_\alpha + {\tilde z}^{({M}-1)}_\alpha &\quad&
\dot{\tilde x}^{({M}-1)} = {\tilde f}^{({M}-1)} + {\tilde w}^{({M}-1)}\\
\tilde v^{({M})}_\alpha  = {\tilde z}^{(M)}_\alpha &\quad& \dot{\tilde x}^{(M)}_\alpha =
{\tilde f}^{(M)}_\alpha + {\tilde w}^{(M)}_\alpha
\end{eqnarray*}
where we have set that
\[\tilde\varphi_\alpha \equiv \tilde v^{(0)}_\alpha \quad{\rm and}\quad
{\tilde g}^{({M+1})}_\alpha\equiv 0.\]
Note that the sensory data $\tilde\varphi_\alpha$ reside at the lowest hierarchical layer and are to
be inferred by the causal states $\tilde v^{(1)}_\alpha$ at the corresponding dynamical orders.
At the highest cortical layer $M$ the causal states $\tilde v^{(M)}_\alpha$ are described by the
spontaneous fluctuations ${\tilde z}^{(M)}_\alpha$ around their means (which have been set to be zero without loss of generality).
Note that the generalized motions of hidden states are still present at the highest cortical level, in just the same way that they manifest at all the other hierarchical levels: the corresponding
spontaneous fluctuations are given by ${\tilde w}^{(M)}_\alpha$.

Separating brain states into causal and hidden states, we can now express the G-density by generalizing
equation~(\ref{generative2}) as
\begin{eqnarray}
p(\tilde\varphi,\tilde\mu) &=& \prod_{\alpha=1}^N p(\tilde\varphi_\alpha,\tilde\mu_\alpha)
= \prod_{\alpha=1}^N p({\tilde\mu_\alpha}^{(M)})\prod_{i=0}^{{M}-1}
p({\tilde\mu_\alpha}^{(i)}|{\tilde\mu_\alpha}^{(i+1)})\nonumber\\
&\Rightarrow& \prod_{\alpha=1}^N p({\tilde x}_\alpha^{(M)},{\tilde v}_\alpha^{(M)})\prod_{i=0}^{{M}-1}
p({\tilde x}_\alpha^{(i)},{\tilde v}_\alpha^{(i)}|{\tilde x}_\alpha^{(i+1)},{\tilde v}_\alpha^{(i+1)})\nonumber\\
&=& \prod_{\alpha=1}^N p({\tilde x}_\alpha^{(M)},{\tilde v}_\alpha^{(M)})\prod_{i=0}^{{M}-1}
p({\tilde x}_\alpha^{(i)}|{\tilde v}_\alpha^{(i)})p({\tilde v}_\alpha^{(i)}|{\tilde
x}^{(i+1)}_\alpha, {\tilde v}^{(i+1)}_\alpha) \label{DHgener2}
\end{eqnarray}
where in the second step we have used ${\tilde\mu_\alpha}^{(i)}=({\tilde x}_\alpha^{(i)},{\tilde v}_\alpha^{(i)})$ and only the causal states ${\tilde v}_\alpha^{(i)}$ are involved in the inter-layer
transitions in the third step.
Also, it must be understood that $p({\tilde x}_\alpha^{(0)}|{\tilde v}_\alpha^{(0)})\equiv 1$ in equation~(\ref{DHgener2}), which appears solely for a mathematical compactness.
The intra-layer conditional probabilities $p({\tilde x}_\alpha^{(i)}|{\tilde v}_\alpha^{(i)})$ are given as
\begin{eqnarray}
p({\tilde x}_\alpha^{(i)}|{\tilde v}_\alpha^{(i)})& =&
p( x^{(i)}_{\alpha[0]},x^{(i)}_{\alpha[1]},\cdots | v^{(i)}_{\alpha[0]},v^{(i)}_{\alpha[1]},
\cdots )\nonumber\\
& = & \prod_{n=0}^\infty p( x^{(i)}_{\alpha[n]}| v^{(i)}_{\alpha[n]}) \label{DHintra1}
\end{eqnarray}
where in the second step we have made use of the assumption of statistical independence among the generalized states at different dynamical orders.
The quantity $p( x^{(i)}_{\alpha[n]}| v^{(i)}_{\alpha[n]})$ specifies the conditional density at the dynamical order $n$ within the hierarchical layer $i$, where the corresponding fluctuations $w^{(i)}_{\alpha[n]}$ are assumed to take Gaussian form as
\begin{equation}\label{DHintra2}
p( x^{(i)}_{\alpha[n]}| v^{(i)}_{\alpha[n]}) \equiv
\frac{1}{\sqrt{2\pi\sigma^{\alpha(i)}_{w[n]}}} \exp \left[ - \left(
x^{(i)}_{\alpha[n+1]}-f^{(i)}_{\alpha[n]}\right)^2 /
\left(2\sigma^{\alpha(i)}_{w[n]}\right)\right].
\end{equation}
The conditional densities $p({\tilde v}_\alpha^{(i)}|{\tilde x}_\alpha^{(i+1)},{\tilde v}_\alpha^{(i+1)})$ appearing in equation~(\ref{DHgener2}) link two successive causal states in the cortical hierarchy which are specified by a similar Gaussian fluctuation for $z_{\alpha[n]}^{(i)}$ via equation~(\ref{inter-layer2}) as
\begin{equation}\label{DHinter2}
p({\tilde v}_\alpha^{(i)}|{\tilde x}_\alpha^{(i+1)},{\tilde v}_\alpha^{(i+1)})
\equiv \prod_{n=0}^\infty \frac{1}{\sqrt{2\pi\sigma^{\alpha(i)}_{z[n]}}} \exp \left[ - \left( v^{(i)}_{\alpha[n]}-g^{(i+1)}_{\alpha[n]}\right) ^2/ \left(2\sigma^{\alpha(i)}_{z[n]}\right)\right].
\end{equation}
What is left unspecified in constructing the G-density fully, i.e. equation~(\ref{DHgener2}), is the prior density $p({\tilde x}_\alpha^{(M)}, {\tilde v}_\alpha^{(M)})$ at the highest cortical layer.
It is given here explicitly as
\begin{eqnarray}\label{DHprior3}
p({\tilde x}_\alpha^{(M)},{\tilde v}_\alpha^{(M)})
&\equiv & \prod_{n=0}^\infty \frac{1} {\sqrt{2\pi\sigma_{w[n]}^{\alpha(M)}}} \exp \left\{ - [x^{(M)}_{\alpha[n+1]}-
f^{(M)}_{\alpha[n]}]^2/\left(2\sigma_{w[n]}^{\alpha(M)}\right)\right\} \nonumber\\
&\times& \prod_{n=0}^\infty \frac{1}{\sqrt{2\pi\sigma_{z[n]}^{\alpha(M)}}} \exp \left\{ - [v^{(M)}_{\alpha[n]}] ^2/\left(2\sigma_{z[n]}^{\alpha(M)}\right)\right\}.
\end{eqnarray}
The prior in the highest cortical layer, equation~(\ref{DHprior3}), comprises
the lateral generalized motions of the hidden states and the spontaneous,
random fluctuations associated with the causal states.
It is assumed that both causal and hidden states fluctuate about zero means.

Next, the Laplace-encoded energy ${E}$ can be written explicitly by substituting equation~(\ref{DHgener2}) into equation~(\ref{minimum-energy}) and incorporating the likelihood and prior densities, equations~(\ref{DHintra2}), (\ref{DHinter2}), and (\ref{DHprior3}), at all hierarchical layers and dynamical orders.
After a straightforward manipulation, we obtain the Laplace-encoded energy for a specific brain variable $\mu_\alpha$ as
\begin{eqnarray*}
E_\alpha(\tilde\mu_\alpha,\tilde\varphi_\alpha) &=&  \sum_{n=0}^\infty \left\{
\frac{1}{2\Omega^{\alpha (M)}_{w[n]}}
\left(x^{(M)}_{\alpha[n+1]}-f^{(M)}_{\alpha[n]}\right)^2 + \frac{1}{2}
\ln\Omega^{\alpha(M)}_{w[n]}\right\}\\
&+& \sum_{n=0}^\infty \left\{ \frac{1}{2\Omega^{\alpha (M)}_{z[n]}}
\left(v^{(M)}_{\alpha[n]}\right)^2 +
\frac{1}{2}\ln\Omega^{\alpha(M)}_{z[n]}\right\} \\
&+& \sum_{i=1}^{{M}-1}\sum_{n=0}^\infty \left\{
\frac{1}{2\Omega^{\alpha(i)}_{w[n]}}
\left(x^{(i)}_{\alpha[n+1]}-f^{(i)}_{\alpha[n]}\right)^2 + \frac{1}{2}
\ln\Omega^{\alpha(i)}_{w[n]} \right\} \\
&+& \sum_{i=0}^{{M}-1}\sum_{n=0}^\infty \left\{
\frac{1}{2\Omega^{\alpha(i)}_{z[n]}}
\left(v^{(i)}_{\alpha[n]}-g^{(i+1)}_{\alpha[n]}\right)^2 +
\frac{1}{2}\ln\Omega^{\alpha(i)}_{z[n]} \right\}
\end{eqnarray*}
where the first and second terms are from prior-densities at the
highest layer, equation~(\ref{DHprior3}), the third term is from
equation~(\ref{DHintra2}), and last term from
equation~(\ref{DHinter2}). A quick inspection reveals that the
first and second terms can be absorbed into the third and fourth
terms, respectively.
Then, the Laplace-encoded energy for multiple brain variables is written compactly as
\begin{eqnarray}\label{DH-FFE1}
E(\tilde\mu,\tilde\varphi) 
&=& \sum_{\alpha=1}^N E_\alpha(\tilde\mu_\alpha,\tilde\varphi_\alpha)\nonumber\\
&=& \frac{1}{2}\sum_{\alpha=1}^N \sum_{n=0}^\infty \sum_{i=1}^{{M}} \left\{\frac{1}{\sigma^{\alpha(i)}_{w[n]}}\left(\varepsilon^{\alpha(i)}_{w[n]} \right)^2 + \ln\sigma^{\alpha(i)}_{w[n]}\right\} \nonumber \\
&+& \frac{1}{2}\sum_{\alpha=1}^N \sum_{n=0}^\infty\sum_{i=0}^{{M}} \left\{\frac{1}{\sigma^{\alpha(i)}_{z[n]}} \left( \varepsilon^{\alpha(i+1)}_{z[n]} \right)^2 + \ln\sigma^{\alpha(i)}_{z[n]} \right\} .
\end{eqnarray}
where we have defined the prediction errors
\begin{eqnarray}
\varepsilon^{\alpha(i)}_{z[n]}  &\equiv&  v^{(i-1)}_{\alpha[n]}-g_{\alpha[n]}^{(i)}
\left(x^{(i)}_{\alpha[n]},v^{(i)}_{\alpha[n]}\right)\label{H-error}\\
\varepsilon^{\alpha(i)}_{w[n]} &\equiv&  x^{(i)}_{\alpha[n+1]} - f_{\alpha[n]}^{(i)}
\left(x^{(i)}_{\alpha[n]},v^{(i)}_{\alpha[n]}\right).\label{D-error}
\end{eqnarray}

Thus, it turns out that the Laplace-encoded energy is expressed essentially as
a sum of the prediction-errors squared and their associated variances.
It appears in equation~(\ref{DH-FFE1}) that the structure of the first term differs
from the second term: In the first term the hierarchical index runs from $i=1$ which
indicates the lowest cortical layer, while the second term includes additional $i=0$ in the
hierarchical sum which designates the sensory data,
$\tilde \varphi\equiv \tilde v^{(0)}$.
Note also in equation~(\ref{H-error}) that
$\varepsilon^{\alpha(M+1)}_{z[n]}=v^{(M)}_{\alpha[n]}$ because the highest
hierarchical layer is at $i=M$, accordingly $g^{(M+1)}_{\alpha[n]}\equiv 0$
by construction.

Table 5 provides the glossary of the mathematical objects involved in the G-density in the full construct for a single brain activity $\mu_\alpha$.

To summarize,  the `full construct' incorporates into the G-density, both multi-layer hierarchies corresponding
to cortical architecture, and multi-scale dynamics in each layer via generalized coordinates. The G-density is expressed as the sequential product of the priors and the likelihoods, cascading down the cortical hierarchy to the
lowest layer where the sensory data are registered (mediated by
causal states), and taking into account the intra-layer
dynamics, mediated by hidden states. The final form of the Laplace-encoded energy,
equation~(\ref{DH-FFE1}), has been derived from equation~(\ref{minimum-energy})
which specifies the Laplace-encoded energy as the (negative) logarithm of the generative
density constructed for the hidden and causal brain states.

\subsection{The full-construct recognition dynamics and neuronal activity}
\label{Recognition-dynamics}

We now describe recognition dynamics incorporating the full construct
(section~\ref{Full-construct}), given the Laplace-encoded energy
${ E}(\tilde\mu,\tilde\varphi)$, equation~(\ref{DH-FFE1}).
In the full construct, the brain states $\tilde \mu_\alpha$
are decomposed into the causal states $\tilde v_\alpha$ which link
the cortical hierarchy and the hidden states $\tilde x_\alpha$ which implement
the dynamical ordering within a cortical layer.

Distinguishing the `path of the modes' from the `modes of the path', see \secref{minimisation},
the learning algorithm for the dynamical causal states on the cortical layer $i$ can be
constructed from
\begin{equation}\label{DH-causal1}
\dot{v}^{(i)}_{\alpha[n]} - D{v}^{(i)}_{\alpha[n]} \equiv -
\kappa_z\hat{v}^{(i)}_{\alpha[n]} \cdot\nabla_{\tilde v_\alpha} {E}
(\tilde\mu,\tilde\varphi)
\end{equation}
where $\kappa_z$ is the learning rate and $\hat{v}^{(i)}_{\alpha[n]}$ is
the unit vector along $v^{(i)}_{\alpha[n]}$.
As mentioned in \secref{minimisation}, the crucial assumption here
is that when the path of modes becomes identical to the modes of
the path, i.e. $\dot{\tilde v}^{(i)}_{\alpha} - D{\tilde v}^{(i)}_{\alpha}\rightarrow
0$, the Laplace-encoded energy ${ E}$ takes its minimum, and \textit{vice
versa}. The gradient operation in the RHS of
equation~(\ref{DH-causal1}) can be made explicit to give
\begin{eqnarray}\label{manipul3}
&&\hat{v}^{(i)}_{\alpha[n]}\cdot\nabla_{\tilde v_\alpha} {E}(\tilde\mu,\tilde\varphi) \nonumber\\
& =& \frac{\partial}{\partial v_{\alpha[n]}^{(i)}} \left[ \frac{1}{2\sigma^{\alpha(i-1)}_{z[n]}}
\left\{\varepsilon^{\alpha(i)}_{z[n]}\right\}^2 +
\frac{1}{2\sigma^{\alpha(i)}_{z[n]}}\left\{\varepsilon^{\alpha(i+1)}_{z[n]}\right\}^2
+ \frac{1}{2\sigma^{\alpha(i)}_{w[n]}}\left\{\varepsilon^{\alpha(i)}_{w[n]}\right\}^2
\right]\nonumber\\
&= &\frac{1}{\sigma^{\alpha(i-1)}_{z[n]}}\varepsilon^{\alpha(i)}_{z[n]}
\frac{\partial \varepsilon^{\alpha(i)}_{z[n]}}{\partial v_{\alpha[n]}^{(i)}}
+\frac{1}{\sigma^{\alpha(i)}_{z[n]}}\varepsilon^{\alpha(i+1)}_{z[n]}\frac{\partial
\varepsilon^{\alpha(i+1)}_{z[n]}}{\partial v_{\alpha[n]}^{(i)}}
+\frac{1}{\sigma^{\alpha(i)}_{w[n]}}\varepsilon^{\alpha(i)}_{w[n]}\frac{\partial
\varepsilon^{\alpha(i)}_{w[n]}}{\partial v_{\alpha[n]}^{(i)}}
\end{eqnarray}
where one can further see that
\[ \frac{\partial \varepsilon^{\alpha(i)}_{z[n]}}{\partial v_{\alpha[n]}^{(i)}}
= - \frac{\partial g^{\alpha(i)}_{z[n]}}{\partial v_{\alpha[n]}^{(i)}},\quad
\frac{\partial \varepsilon^{\alpha(i+1)}_{z[n]}}{\partial v_{\alpha[n]}^{(i)}} = 1,
\quad {\rm and}\quad \frac{\partial
\varepsilon^{\alpha(i)}_{w[n]}}{\partial v_{\alpha[n]}^{(i)}} = -
\frac{\partial f^{(i)}_{\alpha[n]}}{\partial v_{\alpha[n]}^{(i)}}.
\]
The additional auxiliary variables are introduced:
\begin{equation}\label{z-erro-unit}
\xi^{\alpha(i)}_{z[n]} \equiv
\varepsilon^{(i)}_{z[n]}/\sigma^{\alpha(i-1)}_{z[n]} \equiv
\Lambda^{\alpha(i-1)}_{z[n]}\left\{
v^{(i-1)}_{\alpha[n]}-g_{\alpha[n]}^{(i)}\left(x^{(i)}_{\alpha[n]},v^{(i)}_{\alpha[n]}
\right)\right\},
\end{equation}
\begin{equation}\label{w-erro-unit}
\xi^{\alpha(i)}_{w[n]} \equiv
\varepsilon^{(i)}_{w[n]}/\sigma^{(i)}_{w[n]} \equiv
\Lambda^{\alpha(i)}_{w[n]}\left\{ x^{(i)}_{\alpha[n+1]} -
f_{[n]}^{(i)}\left(x^{(i)}_{\alpha[n]},v^{(i)}_{\alpha[n]}\right)\right\},
\end{equation}
where $\Lambda^{\alpha(i)}_{z[n]}$ and $\Lambda^{\alpha(i)}_{w[n]}$ are the
inverse of the variances,
\begin{equation}\label{Precision}
\Lambda^{\alpha(i)}_{z[n]} \equiv 1/\sigma^{\alpha(i)}_{z[n]}\quad{\rm
and}\quad \Lambda^{\alpha(i)}_{w[n]} \equiv 1/\sigma^{\alpha(i)}_{w[n]},
\end{equation}
which are called the \textit{precisions}.
Note that the precisions reflect the magnitude of the prediction errors.

Its is proposed that the auxiliary variables $\xi^{\alpha(i)}_{z[n]}$ and
$\xi^{\alpha(i)}_{w[n]}$ represent \textit{error units}
and that the brain states, $v_{\alpha[n]}^{(i)}$ and $x_{\alpha[n]}^{(i)}$, similarly represent
\textit{state units} or, equivalently,
\textit{representation units}, within neuronal populations
\cite{Friston2009Neural,Friston2010Nature}.

In terms of ‘predictive coding’ or (more generally) hierarchical message
passing in cortical networks\cite{Friston2008One},  equation~(\ref{z-erro-unit}) implies that the error-units
$\xi^{\alpha(i)}_{z[n]}$ receive signals from causal states $v_{\alpha[n]}^{(i-1)}$ lying in immediately lower hierarchical layer and also from
causal and hidden states in the same layer, $v_{\alpha[n]}^{(i)}$ and $x^{(i)}_{\alpha[n]}$, via the generative function $g_{\alpha[n]}^{(i)}$. Similarly, equation~(\ref{w-erro-unit}) implies that the error-units $\xi^{\alpha(i)}_{w[n]}$ specify prediction-error in the
within-layer (lateral) dynamics: $\xi^{\alpha(i)}_{w[n]}$ designates prediction error between the objective
hidden-state $x^{(i)}_{\alpha[n+1]}$ and its estimation from
one-order lower causal- and hidden-states $v_{\alpha[n]}^{(i)}$ and $x_{\alpha[n]}^{(i)}$,
via the different generative function $f_{\alpha[n]}^{(i)}$.

With the help of equation~(\ref{manipul3}), one can recast the learning algorithm equation~(\ref{DH-causal1}) to give the
dynamics of the causal states as
\begin{equation}\label{DH-causal2}
  \dot{v}^{(i)}_{\alpha[n]} = D{v}^{(i)}_{\alpha[n]} + \kappa_z \frac{\partial
g_{[n]}^{\alpha(i)}} {\partial v_{\alpha[n]}^{(i)}}\xi^{\alpha(i)}_{z[n]} - \kappa_z
\xi^{\alpha(i+1)}_{z[n]} + \kappa_z \frac{\partial
f_{\alpha[n]}^{(i)}} {\partial v_{\alpha[n]}^{(i)}}\xi^{\alpha(i)}_{w[n]}
\end{equation}
which shows clearly how hierarchical links are made among nearest-neighbor
cortical layers. Specifically, the representation units of  causal states $v_{\alpha[n]}^{(i)}$ are updated by the error units
$\xi^{\alpha(i+1)}_{z[n]}$ which reside in the layer immediately above,
and also by the error-units $\xi^{\alpha(i)}_{z[n]}$ and $\xi^{\alpha(i)}_{w[n]}$
in the same hierarchical layer, all at the same dynamical order.

The intra-layer dynamics of hidden states are generated similarly as
\begin{eqnarray}
\dot{x}^{(i)}_{\alpha[n]} &\equiv& Dx^{(i)}_{\alpha[n]} - \kappa_w{\hat x}^{(i)}_{\alpha[n]}
\cdot\nabla_{{\tilde x}_\alpha} {E} (\tilde\mu,\tilde\varphi)\nonumber \\
&=& Dx^{(i)}_{\alpha[n]} - \kappa_w \xi^{\alpha(i)}_{w[n-1]} +
\kappa_w \frac{\partial f^{(i)}_{\alpha[n]}}{\partial x^{(i)}_{\alpha[n]}} \xi^{\alpha(i)}_{w[n]}
+ \kappa_w \frac{\partial g^{(i)}_{\alpha[n]}}{\partial x^{(i)}_{\alpha[n]}} \xi^{\alpha(i)}_{z[n]}
\label{DH-hidden2}
\end{eqnarray}
where $\kappa_w$ is the leaning rate. 
In passing to the second line in equation~(\ref{DH-hidden2}), one needs to evaluate
\begin{eqnarray*}
&&{\hat x}^{(i)}_{\alpha[n]}\cdot\nabla_{{\tilde x}_\alpha} {E}(\tilde\mu,\tilde\varphi)\\
\rightarrow &&
\frac{1}{\sigma^{\alpha(i)}_{w[n-1]}}\varepsilon^{\alpha(i)}_{w[n-1]}\frac{\partial
\varepsilon^{\alpha(i)}_{w[n-1]}}{\partial x_{\alpha[n]}^{(i)}} +
\frac{1}{\sigma^{\alpha(i)}_{w[n]}}\varepsilon^{\alpha(i)}_{w[n]}\frac{\partial
\varepsilon^{\alpha(i)}_{w[n]}}{\partial x_{\alpha[n]}^{(i)}}
+ \frac{1}{\sigma^{\alpha(i-1)}_{z[n]}} \varepsilon^{\alpha(i)}_{z[n]}\frac{\partial
\varepsilon^{\alpha(i)}_{z[n]}}{\partial x_{\alpha[n]}^{(i)}},
\end{eqnarray*}
and an explicit evaluation of the derivatives of the prediction errors, equations~(\ref{H-error}) and (\ref{D-error}).
The hidden-state learning algorithm, equation~(\ref{DH-hidden2}), specifies how the representation-units
$x_{\alpha[n]}^{(i)}$ are driven by the error-units in the current layer $i$
at both the immediately lower dynamical order $\xi^{\alpha(i)}_{w[n-1]}$
and the same dynamical order $\xi^{\alpha(i)}_{w[n]}$, and also by the error units
$\xi^{\alpha(i)}_{z[n]}$ in the current layer at the same dynamical order.

To summarize, the hierarchical, dynamical causal structure of
the generative model is fully implemented in the
mathematical constructs given by equations~(\ref{z-erro-unit})
and (\ref{w-erro-unit}) (specifying prediction errors), and
equations~(\ref{DH-causal2}) and (\ref{DH-hidden2}) (specifying
update rules for state-units).

According to these equations, the state units come to encode the
conditional expectations of the environmental causes of sensory
data, and the error units measure the discrepancy between these
expectations and the data. Error units are driven by state units at
the same layer and from the layer below, whereas state units are
driven by error units at the same layer and the layer above. Thus,
prediction errors are passed up the hierarchy (bottom-up) and
predictions (conditional expectations) are passed down the hierarchy
(top-down), fully consistent with predictive coding \cite{rao:1999}.

\begin{table}[b]
\begin{tabular}{p{1.7in}p{4.2in}}

\multicolumn{2}{l} {{\bf Table 6. Mathematical objects for recognition dynamics}} \\
\hline \hline \multicolumn{1}{l} {\bf Symbol} & \multicolumn{1}{l}
{\bf Name \& Description}\\ \hline

&  \\

$\nabla_{\tilde\mu} { E}(\tilde\mu,\tilde\varphi)$ & `Gradient' of the Laplace encoded-energy: Multi-dimensional derivative of the scalar function ${E}$;
which vanishes at an optimum $\tilde\mu^*$. \\

&  \\
\underline{Dynamical construct} & $\dot{\tilde\mu}^{(i)}_{\alpha} - D{\tilde\mu}^{(i)}_{\alpha} = - \kappa_\alpha\nabla_{\tilde \mu_\alpha^{(i)}} {E} (\tilde\mu,\tilde\varphi)$,\quad ${\tilde\mu}^{(i)}_\alpha=({\tilde x}^{(i)}_\alpha,{\tilde v}^{(i)}_\alpha)$ \\
&  \\

$\tilde\mu_\alpha^{(i)}$ &  Generalized brain states: A point in the generalized state space to represent fast `time-dependent' neuronal activity $\mu_\alpha$ on each cortical layer $i$ [see equations~(\ref{inter-layer2}) and (\ref{intra-layer2})]. \\

$\dot {\tilde \mu}_\alpha^{(i)}$, $D{\tilde\mu}_\alpha^{(i)}$ & $\dot{\tilde \mu}_\alpha^{(i)}$ is the `path of the mode'; $D{\tilde \mu}_\alpha^{(i)}$ is the `mode of the path'.  
$\dot{\tilde \mu}_\alpha^{(i)}$ represents the rate of change of a brain state in generalized state space, while $D\tilde\mu_\alpha^{(i)}$ represents the encoded motion in the brain; when the two become identical, i.e. $\dot {\tilde \mu}_\alpha^{(i)} = D\tilde\mu_\alpha^{(i)}$, in the course of recognition dynamics, $E$ reaches its minimum. \\

&  \\
\underline{Static construct} & $\ddot{\mu}_\beta^{(i)} = - \kappa_\beta\hat{\mu}_\beta^{(i)}\cdot\nabla_{\mu_\beta} {E}(\tilde\mu,\tilde\varphi;\theta,\gamma)$, \quad $\mu_\beta=\theta,\gamma$ \\
& \\

$\tilde\Lambda^{\alpha(i)}_z$, $\tilde\Lambda^{\alpha(i)}_w$  & Precisions: Inverse variances in the generalised coordinates [see equation~(\ref{Precision})].\\

$\theta_\alpha^{(i)}$, $\gamma_\alpha^{(i)}$ & Parameters, hyper-parameters: The slow brain states are treated `static' and are associated with $\theta_\alpha^{(i)}$ and $\gamma_\alpha^{(i)}$, respectively, on each cortical layer; where $\theta_\alpha^{(i)}$ appear as
parameters in the generative functions $g_\alpha^{(i)}$ and $f_\alpha^{(i)}$, and $\gamma_\alpha^{(i)}$ are hyper-parameters in the precisions $\Lambda^{\alpha(i)}_z$ and $\Lambda^{\alpha(i)}_w$. \\

$\tilde{\xi}_z^{\alpha(i)}$, $\tilde{\xi}_w^{\alpha(i)}$  & Prediction errors; measuring the discrepancy between
the observation and the evaluation [\textit{e.g.} equations~(\ref{z-erro-unit2}),
(\ref{w-erro-unit2})] \\

\hline

\end{tabular}
\end{table}

\subsection{Parameters and hyper-parameters: Synaptic efficacy and gain}
\label{Parameters}
Thus far we have discussed how environmental variables can be inferred given an appropriate G-density. 
In this section we discuss how the G-density itself can be learned.  It is proposed that  the dynamics of neural systems is captured by three time-scales, $\tau_\mu < \tau_\theta <\tau_\gamma$. The first, $\tau_\mu$, represents the timescale of the dynamics of sufficient statistics of the  encoded in the R-density i,.e  $\mu \equiv (x,v)$ as described above. 
In contrast $\tau_\theta$ and $\tau_\gamma$ represent the slow timescale of synaptic efficacies and gains which are parameterised implicitly in equation~(\ref{DH-FFE1}) through the generative functions, $f$ and $g$, and the variances $\sigma$ (or the precisions $\Lambda$, \eqnref{Precision}), respectively.  Under the FEP slow variables are assumed to be approximately `static' or `time-invariant' in contrast to the `time-varying' neuronal states $\mu$ \cite{Friston2009Neural}.
Second, changes in $\theta$ and $\gamma$ (with respect to a small $\delta t$) have a much
smaller effect on the Laplace-encoded energy (or IFE) than do changes in $\mu$, i.e.
\[
\frac{\partial F}{\partial \theta}\frac{\delta \theta}{\delta t}
 \ll \frac{\partial F}{\partial \mu}\frac{\delta \mu}{\delta t}.
\]
The latter point implies that, from the perspective of
gradient-descent, what is relevant for $\theta$ and
$\gamma$ is not the IFE $F$ but the accumulation, more
precisely the integration of $F$ over time
\cite{Friston2007Synthe}
\begin{equation}\label{Action}
S[F] \equiv \int dt F(\tilde \mu,\tilde \varphi; \theta,\gamma)
\end{equation}
where the time-dependence of $F$ is implicit through the arguments.
To distinguish their different roles, $\theta^{(i)}_\alpha$ are called
\textit{parameters} and $\gamma^{(i)}_\alpha$ are called
\textit{hyper-parameters}, corresponding to brain state $\mu_\alpha$, in each hierarchical layer $i$.
Equations~(\ref{z-erro-unit}) and (\ref{w-erro-unit}) can now be
generalized to include these parameters and hyper-parameters as
\begin{equation}\label{z-erro-unit2}
\xi^{\alpha(i)}_{z[n]} = \Lambda^{\alpha(i-1)}_{z[n]}(\gamma_\alpha^{(i-1)})\left\{
v^{(i-1)}_{\alpha[n]}-g_{\alpha[n]}^{(i)}\left(x^{(i)}_{\alpha[n]},v^{(i)}_{\alpha[n]};\theta_\alpha^{(i)}
\right)\right\},
\end{equation}
\begin{equation}\label{w-erro-unit2}
\xi^{\alpha(i)}_{w[n]} = \Lambda^{\alpha(i)}_{w[n]}(\gamma_\alpha^{(i)}) \left\{ x^{(i)}_{\alpha[n+1]} -
f_{\alpha[n]}^{(i)}\left(x^{(i)}_{\alpha[n]},v^{(i)}_{\alpha[n]};\theta_\alpha^{(i)}\right)\right\}.
\end{equation}
The Laplace-encoded energy including $\theta$ and $\gamma$ may therefore be written as
\begin{eqnarray}\label{DH-FFE2}
E(\tilde\mu,\tilde\varphi;\theta,\gamma)
&=& \frac{1}{2}\sum_{\alpha=1}^N \sum_{n=0}^\infty \sum_{i=1}^{{M}} \left\{ \varepsilon^{\alpha(i)}_{w[n]} \xi^{\alpha(i)}_{w[n]} - \ln\Lambda^{\alpha(i)}_{w[n]}\right\} \nonumber \\
&+& \frac{1}{2}\sum_{\alpha=1}^N \sum_{n=0}^\infty\sum_{i=0}^{{M}} \left\{ \varepsilon^{\alpha(i+1)}_{z[n]} \xi^{\alpha(i+1)}_{z[n]} - \ln\Lambda^{\alpha(i)}_{z[n]} \right\}.
\end{eqnarray}

We are now in a position to write down the recognition dynamics for
the slow synaptic efficacy $\theta$ and for the slower synaptic gain $\gamma$.
Specifically, gradient descent for the parameters $\theta^{(i)}_\alpha$ is
applied using the time-integral of $F$, given in
equation~(\ref{Action}), assuming a static model (i.e., without
dynamical order indices), as
\[
\dot \theta^{(i)}_\alpha = - \kappa_\theta \hat\theta^{(i)}_\alpha
\cdot\nabla_{\theta} S
\]
which, when temporal differentiation is repeated on both sides, gives rise to
\begin{equation}\label{para-grad}
\ddot \theta^{(i)}_\alpha = - \kappa_\theta \hat\theta^{(i)}_\alpha
\cdot\nabla_{\theta} {E}(\tilde\mu,\tilde\varphi;\theta,\gamma).
\end{equation}
After explicitly carrying out the gradient on the RHS of
equation~(\ref{para-grad}), one obtains an equation to minimise  $\theta^{(i)}_\alpha$ corresponding to brain variable $\mu_\alpha$ at cortical layer $i$
\begin{equation}\label{para-grad2}
\ddot \theta^{(i)}_\alpha = \sum_{n=0}^\infty \left[\kappa_\theta \frac{\partial g_{\alpha[n]}^{(i)}}{\partial\theta_\alpha^{(i)}}\xi_{z[n]}^{\alpha(i)}
+ \kappa_\theta \frac{\partial f_{\alpha[n]}^{(i)}}{\partial \theta_\alpha^{(i)}}\xi_{w[n]}^{\alpha(i)}\right]
\end{equation}
where the summation over the dynamic index $n$ reflects the generalized motion over causal as well as hidden states. According to Equation~(\ref{para-grad2}) synaptic efficacy is
influenced by error-units only in the same cortical layer.

Similarly, the learning algorithm for the hyper-parameters $\gamma$, specifically for $\gamma_\alpha^{(i)}$ associated with brain's representation of environmental states $\mu_\alpha$ at cortical layer $i$, is given from
\[
\dot \gamma_\alpha^{(i)} = - \kappa_\gamma \hat\gamma_\alpha^{(i)}
\cdot\nabla_{\gamma} S
\]
which results in
\begin{eqnarray}\label{hyper-grad2}
\ddot \gamma_\alpha^{(i)}
& =& - \frac{1}{2} \sum_{n=0}^\infty\left[\kappa_\gamma \frac{\partial
\Lambda_{w[n]}^{\alpha(i)}}{\partial \gamma_\alpha^{(i)}}\left\{\xi_{w[n]}^{\alpha(i)}\right\}^2
- \kappa_\gamma \frac{\partial}{\partial\gamma_{\alpha}^{(i)}}
\ln\Lambda^{\alpha(i)}_{w[n]}\right] \nonumber\\
&& -\frac{1}{2} \sum_{n=0}^\infty\left[\kappa_\gamma \frac{\partial
\Lambda_{z[n]}^{\alpha(i)}} {\partial \gamma_\alpha^{(i)}}\left\{\xi_{z[n]}^{\alpha(i+1)}\right\}^2 
 - \kappa_\gamma \frac{\partial}{\partial\gamma_\alpha^{(i)}}\ln\Lambda^{\alpha(i)}_{z[n]} \right].
\end{eqnarray}
According to this equation, synaptic gains are influenced by
error units in the same layer $\xi_w^{(i)}$ and also by error units in one-layer above
$\xi_z^{(i+1)}$.

Note that the equations for $\theta$ and $\gamma$,
equations~(\ref{para-grad2}) and (\ref{hyper-grad2}), are \textit{by
construction} second-order differential equations, unlike the
corresponding equations for state-units $\mu$
[equations~(\ref{DH-causal2}) and (\ref{DH-hidden2})], which are
first-order in time \cite{Friston2008NeuroImage}.
Table 6 provides the summary of mathematical symbols appearing in the recognition dynamics in the dynamical construct and also in the static construct.

To summarize the FEP prescribes \textit{recognition dynamics} by  gradient descent with respect to the sufficient statistics ${\tilde\mu}$, parameters $\theta$, and hyper-parameters $\gamma$
on the Laplace-encoded energy ${E}({\tilde\mu},\tilde\varphi;\theta,\gamma)$, given the sensory input $\tilde\varphi$.
At the end of this process, an optimal $\tilde\mu^*$ is specified which represents the brain's posterior expectation of the environmental cause of the observed sensory data.
In theory the second term in the IFE $F$, \eqnref{FE4-5}, can be fixed according to
\eqnref{gammastar} thereby completing the minimization of the IFE, although in practice this is rarely done and the focus is on approximating the means, parameters and hyper-parameters.

This whole minimization process is expressed abstractly as
\begin{equation}\label{argminmu}
{\tilde\mu}^* = {\rm arg}\ \min_{\tilde\mu} F(\tilde\mu,\tilde\varphi)
\end{equation}
where $\tilde\mu^*$ is the minimizing (optimal) solution
and the conditional dependence $m$ is expressed
explicitly. 
The resulting minimized IFE can be calculated by
substituting the optimizing $\tilde\mu^*$ for $\tilde\mu$ as
\[ F^* = F({\tilde\mu}^*,\tilde\varphi~).\]

The only remaining task is to specify the generative functions $f$ and $g$, which will depend on the particular system being modelled.
We have utilised a concrete model in our calculation in \secref{action}.
Examples of various generating functions have already been provided\cite{Friston2008One,Friston2010BC,friston2016active, pezzulo2015active,pio2016active}, to which we refer the reader.

\subsection{Active inference on the full construct}
The IFE also accounts for an active inference by minimising the IFE with respect to action, for which a formal procedure can be written as
\begin{equation}\label{argminaction}
a^* = {\rm arg}\ \min_a F(\tilde\mu,\tilde\varphi(a))
\end{equation}
where $a^*$ is the minimizing solution.
Similarly with equation~(\ref{grad-action0}) we can write down the gradient descent scheme for the minimisation in the full construct for action corresponding to brain's representation $\mu_\alpha$ as 
\begin{equation}\label{grad-action2}
\dot a_\alpha  = -\kappa_a \hat{a}_\alpha\cdot\nabla_{a_\alpha} E(\tilde\mu,\tilde\varphi(a))
\end{equation}
where \eqnref{DH-FFE2} is to be used for the Laplace-encoded energy.
Then, after the gradient operation is completed, the organism's action is implemented explicitly in the brain as
\begin{equation}
\dot a_\alpha= -\kappa_a\sum_{n=0}^\infty \frac{d\tilde\varphi_{\alpha[n]}}{da_\alpha}\Lambda^{\alpha(0)}_{z[n]} \varepsilon_{z[n]}^{\alpha(1)}
\end{equation}
where $\varepsilon_{z[n]}^{\alpha(1)} = \varphi_{\alpha[n]}- g_{\alpha[n]}^{(1)}(x_{\alpha[n]}^{(1)},v_{\alpha[n]}^{(1)}; \theta^{(1)}_\alpha)$ is the  prediction-error associated with learning of the sensory data on the dynamical order $n$ at the lowest cortical layer and $\Lambda^{\alpha(0)}_{z[n]}=\Lambda^{\alpha(0)}_{z[n]}(\gamma_\alpha^{(0)})$ is the precision of the sensory noise. To our knowledge most existing models of active inference under the  FEP require one to provide an explicit world model.  Thus  an important goal for for future  work will be to develop agent based models that work with the full construct.

\section{Discussion}
The FEP framework is an ambitious project, spanning a chain of reasoning from fundamental principles of biological self-maintenance essential for sustainable life, to a mechanistic brain theory that proposes to account for a startling range of properties of perception, cognition, action and learning. It draws conclusions about neurocognitive mechanisms from extremely general statistical considerations regarding the viability of organism's survival in unpredictable environments. Under certain assumptions - which we discuss in more detail below - it entails a hierarchical predictive processing model geared towards the inference and control of the hidden causes of sensory inputs, which both sheds new light on existing data about functional neuroanatomy and motivates a number of specific hypotheses regarding brain function in health and in  disease. At the same time, the current status of much of the research under the rubric of the FEP does depend on, to different degrees, a variety of assumptions and approximations, both at the level of the overarching theory and with regard to the specific implementation (or ‘process theory’) the theory proposes. In this section, we discuss the consequences of some of more important of these assumptions and approximations, with respect to the framework and implementation described in the body of this paper.

A central assumption in this (representative) exposition of the FEP is that the brain utilizes properties of Gaussian distributions in order to carry out probabilistic computation. Specifically, the Laplace approximation assumes a Gaussian functional form for the R-density and G-density which are encoded by sufficient statistics, see \secref{laplaceencoding}. Additionally, it is assumed that the R-density is tightly peaked, i.e., the variance and covariance are small, see \secref{laplaceencoding}. This assumption implies that an organism only represents the expectation value of environmental variables, and not a distribution over states, see \cite{bogacz2015tutorial,pouget2013probabilistic,knill2004bayesian} for a nice descriptions of this assumption. At first glance this assumption may appear troublesome, because it suggests that organisms do not directly represent the uncertainty of environmental variables (hidden causes of sensory signals). This worry is misplaced, however, since representations of uncertainty enter into the FEP formalism via precisions on the expectations of brain states that comprise the G-density, see \eqnref{multi}. Intuitively this means that organisms do not encode uncertainty about world states per se, but rather
uncertainties about their model of how hidden causes relate to each other and to sensory signals.  

The main advantage of adopting Gaussian assumptions is that they vastly simplify the implementation of the FEP, and make it formally equivalent to the more widely known predictive coding framework \cite{elias1955predictive,clark2013whatever,friston2009predictive}, see the Introduction. Furthermore, it can be argued this implementation is compatible with  a plausible neuronal functional architecture in terms of message passing in cortical hierarchies \cite{Friston2005Phil}. Specifically, inferred variables (hidden causes) can be represented in terms of neural firing rates; the details of generative models encoded as patterns of synaptic connectivity, and the process of IFE minimisation by the relaxation of neuronal dynamics \cite{bastos2012canonical}. The concept of hierarchical generative models, see \secref{learning}, also maps neatly onto the hierarchical structure of cortical networks, at least in the most frequently studied perceptual modalities like vision. Here, the simple idea is that top-down cortical signalling conveys predictions while bottom-up activity returns prediction errors \cite{bastos2012canonical}. However, it remains an open question whether representing the world in terms of Gaussian distributions is sufficient given the complexities of real-world sensorimotor interactions. For example, standard robotics architectures have long utilized practical strategies for representing more complex distributions \cite{thrun2005probabilistic} including (for example) multimodal peaks \cite{otworowska2014counter}.Other authors have proposed that brains engage in Bayesian sampling rather than the encoding of probability distributions, suggesting that sampling schemes parsimoniously explain classic cognitive reasoning errors \cite{knill2004bayesian}.  Whether these alternate schemes can be used to construct more versatile and behaviourally powerful implementations of the FEP, and whether they remain compatible with neuronally plausible process theories, remains to be seen. 

The minimisation of IFE, for both inference and learning, is assumed to be implemented as a gradient descent scheme. While this has the major advantage of transforming difficult or infeasible inference problems into relatively straightforward optimization problems, it is not clear whether the propose gradient descent schemes always have good convergence properties. For example, the conditions under which gradient descent will become stuck in local minima, or fail to converge in an appropriate amount of time, are not well understood. Furthermore, parameters such as learning rate will be crucial for the timely inference of the dynamics of variables, as well as central to the dynamics of control, see \figref{fig1} and \ref{fig2}. Parameters like these, which play important roles in the estimation of – but not specification of – the IFE, can be incorporated into process theories in many ways, with as yet no clear consensus (though see, for one proposal \cite{joffily2013emotional}).
   
The implementation described in this paper supports inference in dynamical environments. This is based on the concept of generalised motions, whereby it is assumed that the brain infers not only the current value of environmental variables (e.g.,position) but also their higher-order derivatives (i.e., velocity, acceleration, jerk, etc.). This involve  both that the relevant sensory noise is differentiable, and, that interactions between derivatives are linear \cite{Friston2006NeuroImage}. The extent to which these assumptions are justifiable remains unclear, as does the utility of encoding generalized motions in practical applications. It is likely, for example, that signal magnitudes after the second derivative will be small and carry considerable noise, thus practical usefulness of including  higher order derivatives is unclear, although this may be justifiable in some cases \cite{balaji2011bayesian}.

Under active inference, prediction errors are minimised by acting on the world to change sensory input, rather than by modifying predictions. Active inference therefore depends on the ability to make conditional predictions about the sensory consequences of actions. 
To achieve this the FEP assumes that agents have a model of the relationship between action and sensation, in the form of an inverse model, in addition to their generative model \cite{seth2014predictive,friston2016active}. In the general case the specification of an inverse model is non-trivial \cite{wolpert1997computational}, which at first glance seems like a strong assumption. However, the  FEP suggests generation of motor actions are driven through the fulfilment of proprioceptive predictions only, where relations between actions and (proprioceptive) sensations are assumed to be relatively simple such that minimisation of prediction error can be satisfied by simple reflex arcs \cite{Friston2010BC,friston2011optimal}. On this view, action only indirectly affects exteroceptive or interoceptive sensations,  obviating the need for complicated inverse models like those  described in the motor control literature \cite{wolpert1997computational,friston2011optimal}. In the implementation of the FEP given in this paper there is no distinction between different types of sensory input.

In \secref{action} we showed that behaviour is extremely sensitive to precisions. This is often presented as an advantage of the framework, allowing an agent to balance sensory inputs against internal predictions in an optimal and context sensitive manner, through precision weighting (which is associated with attention) \cite{clark2013whatever}. Supposedly the appropriate  regulation of precision should  also emerge as a consequence of the minimisation of free energy, see \secref{learning} for a description of this.  But how the interplay between brain states and precisions will unfold in an active agent involved in a complex behaviour is far from clear.

Where do the priors come from?  This is an intuitive way to put a key challenge for models involving Bayesian inference  \cite{kass:1989}. To some extent the FEP circumvents this problem via the concept of hierarchical models, which maps neatly onto the framework of `empirical Bayes' \cite{Casella:2002}. In this view, the hierarchical structure allows priors at one level to be supplied by posteriors at a higher level. Sensory data are assumed to reside only at the lowest level in the hierarchy, and the highest level is assumed to generate only spontaneous random fluctuations. While this is a powerful idea within formal frameworks, its practicality for guiding inference in active agents remains to be established. 

These discussion points merely scratch the surface of the promises and pitfalls of the FEP formalism, a formalism which is rapidly advancing both in its theoretical aspects and in its various implementations and applications.  Nevertheless, research directed towards addressing these issues should further clarify both the explanatory power and the practical utility of this increasingly influential framework.  In this paper, we have focused on encapsulating within a single presentation the essential mathematical aspects of the FEP and its implementation. In doing so we hope to clarify the scientific contributions of the FEP, facilitate discussions of some of the core issues and assumptions underlying it, and motivate additional research to explore how far the grand ambitions of the FEP can be realized in scientific practice.

\label{discussion}
\section{Acknowledgements}
The work of C.S.K. was supported by a special fund granted by the
Chonnam National University. C.S.K. is grateful to the
hospitality of the School of Engineering aand Informatics at the University of Sussex where he spent a sabbatical. Support is also gratefully
acknowledged from the Dr. Mortimer and Theresa Sackler Foundation.

\appendix
\section{Variational Bayes: Ensemble learning}
\label{EnsembleLearning}
Here, we present an alternative approach to how the brain may achieve true posterior density, which makes no assumptions about how the R-density is encoded in the brain's state; namely the Laplace approximation for the R-density is dispensed with.
Technically the above method is termed `Generalized Filtering' in \cite{friston2010generalised} and the present one `Variational Filtering' in \cite{Friston2008NeuroImageTECH}.

According to equation~(\ref{FFE}) the IFE is a functional of the R-density $q(\vartheta)$ where the variable $\vartheta$ denotes the environmental states collectively.
The environmental sub-states $\vartheta_\alpha$, $\alpha=1,2,\cdots, N$, must vary on distinctive time-scale, $\tau_1 < \tau_2 < \cdots < \tau_N$, where $\tau_\alpha$ is associated with $\vartheta_\alpha$, conforming to physics laws, in general.
Then, the sub-densities may be assumed to be statistically-independent to allow the \textit{factorization approximation} for $q(\vartheta)$ as
\begin{equation}\label{meanfield}
q(\vartheta)\equiv \prod_{i=1}^N q_\alpha(\vartheta_\alpha).
\end{equation}
Equation~(\ref{norm1}) gives rise to the individual normalization
condition:
\[
\int d\vartheta~q(\vartheta)=\prod_{\alpha=1}^N\int d\vartheta_\alpha~q_\alpha(\vartheta_\alpha)= 1
\]
which asserts that
\begin{equation}\label{normi}
\int d\vartheta_\alpha~q_\alpha(\vartheta_\alpha)= 1.
\end{equation}
When the factorization approximation, equation~(\ref{meanfield}) is substituted into equation~(\ref{FFE}), the IFE is written as
\begin{eqnarray*}
F& =&
\int\prod_{\alpha}\left[d\vartheta_\alpha q_\alpha(\vartheta_\alpha)\right]\left\{
E(\vartheta,\varphi) + \sum_ \sigma\ln q_\sigma(\vartheta_\sigma)\right\}\\
&\equiv& F[q(\vartheta);\varphi]
\end{eqnarray*}
where the last expression indicates explicitly that the IFE is to be treated as a \textit{functional} of the R-density. 
We now optimize the IFE functional by taking the variation of $F$ with respect to a particular R-density $q_\beta(\vartheta_\beta)$. We treat the remainder of the ensemble densities as constant and use the normalization constraint, equation~(\ref{norm1}), in the form
\begin{equation}\label{Lmultiple}
\lambda\left( \prod_\alpha\int d\vartheta_\alpha~q_\alpha(\vartheta_\alpha) - 1\right)=0 
\end{equation}
where $\lambda$ is a Lagrange multiplier.

A straightforward manipulation brings about
\[
\delta_\beta F=\int d\vartheta_\beta\left\{\int \prod_{\alpha\neq \beta} d\vartheta_\alpha q_\alpha(\vartheta_\alpha)\left(E(\vartheta,\varphi) + \sum_\sigma \ln q_\sigma(\vartheta_\sigma)\right) + 1 +\lambda \right\}\delta q_\beta
\]
where $\delta_\beta$ represents a functional derivative with respect to $q_\beta(\vartheta_\beta)$.
Next, by imposing $\delta_{\beta} F\equiv 0$ it follows that the integration must vanish identically for any change in $\delta q_\beta$,
\[
\int \prod_{\alpha\neq \beta} d\vartheta_\alpha
q_\alpha(\vartheta_\alpha)\left(E(\vartheta,\varphi) + \sum_\sigma\ln
q_\sigma(\vartheta_\sigma)\right) + 1 +\lambda = 0
\]
which is to be solved for $q_\beta(\vartheta_\beta)$.
The result brings out the optimal density for the sub-state $\vartheta_\beta$ as
\begin{equation}\label{recog0}
q^*_\beta = \exp\left\{ -(\lambda + 1) - \sum_{\sigma\neq \beta} \int
\prod_{\alpha\neq\beta}d\vartheta_\alpha q_\alpha(\vartheta_\alpha)\ln q_\sigma(\vartheta_\sigma) - {\cal
E}_\beta(\vartheta_\beta,\varphi)\right\}
\end{equation}
where use has been made of the definition
\begin{equation}\label{varenergy}
{\cal E}_\beta(\vartheta_\beta,\varphi) \equiv \int\prod_{\alpha\neq \beta}
d\vartheta_{\alpha}q_\alpha(\vartheta_\alpha)E(\vartheta,\varphi)
\end{equation}
which is the partially-averaged energy \cite{Friston2006Physio,Friston2006NeuroImage}.
Here, it is worthwhile to note that the following relation holds
\[
\int d\vartheta_{\beta}~q_\beta(\vartheta_\beta){\cal E}_\beta(\vartheta_\beta,\varphi)
= \int d\vartheta~ q(\vartheta)E(\vartheta,\varphi),
\]
which states that the expectation of the partially-averaged energy
${\cal E}_\beta(\vartheta_\beta,\varphi)$ under $q_\beta(\vartheta_\beta)$ is the
average energy, i.e. the first term in equation~(\ref{Helmholtz}).
The undetermined Lagrange multiplier is now fixed by the
normalization constraint, equation~(\ref{normi}), which results in
\[
\left[\int d\vartheta_{\beta}~e^{-{\cal E}_\beta(\vartheta_\beta,\varphi)}\right]
\exp\left\{ -(\lambda + 1) - \sum_{\sigma\neq \beta}\int \prod_{\alpha\neq{\beta}}d\vartheta_{\alpha} q_\alpha(\vartheta_\alpha) \ln q_\sigma(\vartheta_\sigma)\right\} = 1,
\]
which is to be solved for $\lambda$.
When the \textit{determined} $\lambda$ is substituted back into equation~(\ref{recog0}), the
resulting \textit{ensemble-learned} R-density can be expressed formally as \footnote{Note that the minus sign arises in
the exponent because we have defined the energy as
equation~(\ref{energy}) differently from other papers on the free energy principle. We have made this
choice because our definition resembles the Boltzmann factor in the
canonical ensemble in statistical physics.}
\begin{equation}\label{canonical}
q^*_\beta(\vartheta_\beta) = \frac{1}{Z_\beta}e^{-{\cal E}_\beta(\vartheta_\beta,\varphi)}
\end{equation}
where $Z_\beta$ has been defined to be
\begin{equation}\label{partition}
Z_\beta \equiv \int d\vartheta_\beta~e^{-{\cal E}_\beta(\vartheta_\beta,\varphi)}.
\end{equation}
The superscript $*$ appearing in $q^*_\beta$ indicates that it is the solution which optimizes the IFE.
The functional form of equation~(\ref{canonical}) is reminiscent of the equilibrium canonical ensemble in statistical physics in which the normalization factor $Z_\beta$ is called the \textit{partition function} of the subsystem $\{\vartheta_\beta\}$ \cite{SM}.

Under the factorization approximation, by substituting equation~(\ref{canonical}) into equation~(\ref{meanfield}), the R-density becomes
\begin{equation}\label{equilibrium}
q^*(\vartheta) = \frac{1}{Z_T}e^{-{\cal E}_T(\vartheta,\varphi)}
\end{equation}
where
\[ {\cal E}_T(\vartheta,\varphi) \equiv \sum_{\alpha=1}^N~{\cal E}_\alpha
(\vartheta_\alpha,\varphi) \quad {\rm and} \quad Z_T \equiv \prod_{\alpha=1}^N
Z_\alpha = \int d\vartheta e^{-{\cal E}_T(\vartheta,\varphi)}.
\]
In equation~(\ref{equilibrium}) $Z_T$ may be called the `total' partition function of the environmental
states and ${\cal E}_T$ is the sum of the partially-averaged energies. Note that, as a consequence of the ensemble-learning, the optimizing R-density approximates the posterior density $p(\vartheta|\varphi)$ (see \secref{IFE} and below).
In principle, the optimizing R-density, equation~(\ref{equilibrium}), completes the ensemble-learning of the sensory data.
However, it does not provide a functionally fixed-form for the optimal R-density.
This is because the partially-averaged energy appearing on the RHS of equation~(\ref{equilibrium}) is a functional of the R-density itself (see equation~(\ref{varenergy})).
One possible way to obtain a closed form of $q^*(\vartheta,\varphi)$ is to seek a \textit{self-consistent solution}:
One starts with an educated guess (an `ansatz') for the optimal R-density to evaluate the partially-averaged energy, equation~(\ref{varenergy}) and uses the outcome to update the R-density, equation~(\ref{canonical}).
This iterative process is to be continued until a convergence reaches between estimation and evaluation of the R-densities.

We now exploit what actually the optimal R-density, $q^*_\beta(\vartheta_\beta)$ given in equation~(\ref{canonical}), is.
The partially averaged-energy appearing in $q^*_\beta$ can be manipulated as
\begin{eqnarray}\label{opR1}
{\cal E}_\beta(\vartheta_\beta,\varphi) &=& \int\prod_{\alpha\neq \beta}
d\vartheta_{\alpha}q_\alpha(\vartheta_\alpha)E(\vartheta,\varphi) \nonumber\\
&=& - \sum_\sigma\int\prod_{\alpha\neq \beta} d\vartheta_{\alpha}q_\alpha(\vartheta_\alpha)\ln p(\vartheta_\sigma,\varphi_\sigma),
\end{eqnarray}
where we have used the factorization approximation for the G-density appearing in the energy $E=-\ln p(\vartheta,\varphi)$ as
\begin{equation}
p(\vartheta,\varphi) = \prod_\sigma p(\vartheta_\sigma,\varphi_\sigma) = \prod_\sigma p(\vartheta_\sigma|\varphi_\sigma)p(\varphi_\sigma).
\end{equation}
Next, one can separate out the environmental sub-state $\vartheta_\beta$ among summation on the RHS of equation~(\ref{opR1}) to cast it into
\begin{equation}
{\cal E}_\beta(\vartheta_\beta,\varphi) = - \ln p(\vartheta_\beta,\varphi_\beta)
- \sum_{\sigma\neq \beta}\int\prod_{\alpha\neq \beta} d\vartheta_{\alpha}q_\alpha(\vartheta_\alpha) \ln p(\vartheta_\sigma,\varphi_\sigma).
\end{equation}
Then, it follows from equation~(\ref{canonical}) that
\[
q^*_\beta(\vartheta_\beta) = \frac {e^{-{\cal E}_\beta(\vartheta_\beta,\varphi)}}{\int d\vartheta_\beta~e^{-{\cal E}_\beta(\vartheta_\beta,\varphi)}}
\rightarrow \frac{p(\vartheta_\beta,\varphi_\beta)}{\int d\vartheta_\beta p(\vartheta_\beta,\varphi_\beta)}
= p(\vartheta_\beta|\varphi_\beta),
\]
where the last step can be obtained by noticing the identity,
$p(\vartheta_\beta,\varphi_\beta)=p(\vartheta_\beta|\varphi_\beta)p(\varphi_\beta)$, and
$ \int d\vartheta_\beta p(\vartheta_\beta,\varphi_\beta)=p(\varphi_\beta)$.
Finally, the ensemble-learned R-density, equation~(\ref{equilibrium}), is given by
\begin{equation} \label{RdirPost}
q^*(\vartheta) = \prod_\alpha q_\alpha^*(\vartheta_\alpha) = \prod_\alpha p(\vartheta_\alpha|\varphi_\alpha) = p(\vartheta|\varphi).
\end{equation}
Equation~(\ref{RdirPost}) states that the R-density $q(\vartheta)$ is \textit{directed} to the posterior
$p(\vartheta|\varphi)$ when the IFE, equation~(\ref{FFE}) is minimized, conforming to the idea of variational Bayes.

By substituting the optimal R-density, equation~(\ref{RdirPost}), into expression for IFE given in equation~(\ref{FFE}), we can also obtain the minimized IFE as
\begin{eqnarray}\label{microFE}
F^* &=& \int d\vartheta~ q^*(\vartheta)\ln \frac{q^*(\vartheta)} {p(\vartheta,\varphi)} \nonumber\\
&=& \int d\vartheta~ q^*(\vartheta)\ln \frac {p(\vartheta|\varphi)}{p(\vartheta|\varphi)p(\varphi)} \nonumber\\
&=& - \ln p(\varphi)\int d\vartheta~ q^*(\vartheta)\nonumber \\
&=&  -\ln p(\varphi).
\end{eqnarray}
where we have used equation~(\ref{RdirPost}) in moving to second line and the normalization condition for $q^*(\vartheta)$ in the last step.
Note that we have made it explicit that the sensory density $p(\varphi)$ is conditioned on the biological agent $m$.
Thus, we have come to a conclusion that the minimum IFE provides a \textit{tight bound} on surprisal.

In summary, the variation of the IFE functional with respect to the R-density (ensemble-learning) has allowed us to specify an optimal (ensemble-learned) R-density, $q^*(\vartheta,\varphi)$, selected among an ensemble of
R-densities.
The specified R-density is the brain's solution to statistical inference of the posterior density about the environmental states given sensory inputs.
The minimum IFE, fixed in this way, is identical to the surprisal.
To fulfill this it was assumed that distinctive independent time-scales characterize environmental sub-states (the factorization approximation).
The ensemble-learned R-density of each partitioned variable set $\vartheta_\beta$, $q_\beta^*(\vartheta_\beta)$, is specified by the corresponding partially-averaged energy (see equation~(\ref{canonical})).
The influence from other environmental sets $\{\vartheta_\sigma\}$ ($\sigma\neq \beta$)
occurs as their average effect: Their complicated interactions have been averaged out in equation~(\ref{varenergy}).
In this sense, $\vartheta_\beta$ may be regarded as a `mean-field' of the environmental states.
Accordingly, the procedure described in the above is sometimes referred to as a \textit{mean-field approximation} \cite{Friston2008NeuroImage,Friston2008NeuroImageTECH}.

\section{Dynamic Bayesian Thermostat}
\label{code}

\begin{lstlisting}
% A Simple Bayesian Thermostat
% A free energy principle for action and perception sciences: A mathematical evaluation
% Christopher L. Buckley,Chang Sub Kim, Simon M. McGregor and Anil K. Seth
clear;
rng(6);
%simulation params
simTime=100; dt=0.005; time =0:dt:simTime;
N =length(time);
action =true;
%Generetaive Model Parameters
Td = 4; %desired temperature

%Time we trun on action
actionTime =simTime/4;

%initialise sensors
rho_0(1) =0;
rho_1(1)=0;

%sensory variances
Omega_z0 =0.1;
Omega_z1 =0.1;
%hidden state variances
Omega_w0 =.1;
Omega_w1 =.1;

%Params for generative process
T0 = 100; %temperature at x=0

%intialise brain state variables
mu_0(1)=0;
mu_1(1)=0;
mu_2(1)=0;

% sensory noise in the negerateive proess
zgp_0 = randn(1,N)*.1;
zgp_1 = randn(1,N)*.1;

%Initialise action varaible
a(1) =0;

%Initialise generative process
x_dot(1) = a(1);
x(1) = 2;
T(1) = T0/(x(1)^2+1);
Tx(1)= -2*T0*x(1)*(x(1)^2+1)^-2;
T_dot(1) =  Tx(1)*(x_dot(1));

%Initialise sensory input
rho_0(1) =  T(1);
rho_1(1) =  T_dot(1);

%Intialise error terms
epsilon_z_0 = (rho_0(1)-mu_0(1));
epsilon_z_1 = (rho_1(1)-mu_1(1));

epsilon_w_0 = (mu_1(1)+mu_0(1)-Td);
epsilon_w_1 = (mu_2(1)+mu_1(1));

%Intialise Variational Energy
IFE(1) = 1/Omega_z0*epsilon_z_0^2/2 ...
    + 1/Omega_z1*epsilon_z_1^2/2 ...
    +1/Omega_w0*epsilon_w_0^2/2 ...
    +1/Omega_w1*epsilon_w_1^2/2 ...
    +1/2*log(Omega_w0*Omega_w1*Omega_z0*Omega_z1);


%Gradient descent learning params
k=.1; %for inference
ka=.01; %for learning

for i=2:N
    
    %The generative process
    x_dot(i) = a(i-1);%action
    x(i) = x(i-1)+dt*(x_dot(i));
    T(i) = T0/(x(i)^2+1);
    Tx(i)= -2*T0*x(i)*(x(i)^2+1)^-2;
    T_dot(i) =  Tx(i)*(x_dot(i));
    
    rho_0(i) =  T(i) + zgp_0(i); %calclaute sensory input
    rho_1(i) =  T_dot(i) + zgp_1(i);
    
    %The generative model
    epsilon_z_0 = (rho_0(i-1)-mu_0(i-1));%error terms
    epsilon_z_1 = (rho_1(i-1)-mu_1(i-1));
    
    epsilon_w_0 = (mu_1(i-1)+mu_0(i-1)-Td);
    epsilon_w_1 = (mu_2(i-1)+mu_1(i-1));
    
    IFE(i) = 1/Omega_z0*epsilon_z_0^2/2 ...
        +1/Omega_z1*epsilon_z_1^2/2 ...
        +1/Omega_w0*epsilon_w_0^2/2 ...
        +1/Omega_w1*epsilon_w_1^2/2 ...
        +1/2*log(Omega_w0*Omega_w1*Omega_z0*Omega_z1);
    
    mu_0(i)  = mu_0(i-1) ...
        +dt*(mu_1(i-1)-k*(-epsilon_z_0/Omega_z0 ...
        +epsilon_w_0/Omega_w0));
    
    mu_1(i)  = mu_1(i-1) +dt*(mu_2(i-1)- k*(-epsilon_z_1/Omega_z1 ...
        +epsilon_w_0/Omega_w0+epsilon_w_1/Omega_w1));
    
    mu_2(i)  = mu_2(i-1)...
        +dt*-k*(epsilon_w_1/Omega_w1);
    
    
    if(time(i) >25)
        a(i) = a(i-1) +dt*-ka*Tx(i)*epsilon_z_1/Omega_z1; %active inference
    else
        a(i) = 0;
    end
end
figure(1);clf;



figure(1);

subplot(5,1,1)
plot(time,T); hold on;
plot(time,x); hold on;
legend('T','x')

subplot(5,1,2)
plot(time,mu_0,'k');hold on;
plot(time,mu_1,'m');hold on;
plot(time,mu_2,'b');hold on;

legend('\mu','\mu','\mu');

subplot(5,1,3)
plot(time,rho_0,'k');hold on;
plot(time,rho_1,'m');hold on;

legend('\rho','\rho');

subplot(5,1,4)
plot(time, a,'k');
ylabel('a')

subplot(5,1,5)
plot(time, IFE,'k'); xlabel('time');hold on;
ylabel('IFE')

\end{lstlisting}




\end{document}